\documentclass[onecolumn,a4paper,12pt,oneside]{article} 
\usepackage[top=2.5cm,left=2.5cm,right=2.5cm,bottom=3cm]{geometry}    
\usepackage[nostamp]{draftwatermark} 
\SetWatermarkText{\bf DRAFT}
\SetWatermarkAngle{45}
\SetWatermarkScale{4}
\SetWatermarkLightness{0.85}     
\usepackage[parfill]{parskip}    
\usepackage{graphicx}
\usepackage{amssymb}
\usepackage{amsmath}
\usepackage{amsfonts}
\usepackage{amsthm}
\usepackage{epstopdf}
\usepackage[usenames,dvipsnames,svgnames,table]{xcolor}
\usepackage{tikz}
\usetikzlibrary{patterns}
\usepgflibrary{patterns}
\usepackage{pgfplots}
\usepgfplotslibrary{polar}
\usepgfplotslibrary{fillbetween}
\usepackage{array}
\usepackage[T1]{fontenc}
\usepackage[utf8]{inputenc}
\usepackage{booktabs} 
\usepackage{array} 
\usepackage{paralist} 
\usepackage{listings} 
\usepackage{verbatim} 
\usepackage{subfig} 
\usepackage[colorlinks=true]{hyperref}
\hypersetup{colorlinks=true, linktocpage=true, linkcolor=Blue, citecolor=Green,
hypertexnames=true, urlcolor=Blue,pdftex}
\usepackage[english]{babel}
\usepackage{wrapfig}
\usepackage{multirow}
\usepackage{quoting}
\usepackage{rotating}
\quotingsetup{font=normalsize}
\theoremstyle{plain}

\usepackage{bm}
\usepackage{abstract}
\usepackage{cite}
\usepackage{float}
\usepackage{textcomp} 	
\usepackage{tabularx} 
\usepackage{caption}
\usepackage{authblk} 
\usepackage{eso-pic} 
\usepackage{lineno} 
\providecommand{\keywords}[1]{\textbf{{Key words: }} #1} 

\pgfplotsset{compat=1.5.1}

\newcommand{\be}{\begin{equation}}
\newcommand{\ee}{\end{equation}}
\newcommand{\bsp}{\begin{split}}
\newcommand{\esp}{\end{split}}

\renewcommand{\Phi}{\varPhi}

\newcommand{\eps}{\varepsilon}

\newcommand{\Sy}{\mathbb{S}}

\renewcommand{\Theta}{\varTheta}
\renewcommand{\Psi}{\varPsi}
\renewcommand{\Sigma}{\varSigma}
\newcommand{\A}{\mathbb{A}}
\newcommand{\B}{\mathbb{B}}

\newcommand{\D}{\mathbb{D}}
\newcommand{\Q}{\mathbb{Q}}
\renewcommand{\Delta}{\varDelta}
\renewcommand{\phi}{\varphi}
\renewcommand{\psi}{\varPsi}

\newcommand{\nud}{\nu_{12}}
\newcommand{\bs}{\begin{split}}
\newcommand{\es}{\end{split}}
\newcommand{\e}{\mathrm{e}}
\renewcommand{\i}{\mathrm{i}}
\newcommand{\wE}{\widehat{E}}
\newcommand{\wG}{\widehat{G}}
\newcommand{\wnu}{\widehat{\nu}}
\newcommand{\vf}{\textsl{v}_{\textsl{f}}}

\title{{\LARGE{\bf Auxetic laminates composed of  plies with special orthotropy}}\bigskip\\{\small - PREPRINT-\smallskip\\Final version: \href{https://doi.org/10.1016/j.euromechsol.2025.105766}{https://doi.org/10.1016/j.euromechsol.2025.105766}}}
\author{Paolo VANNUCCI\bigskip\\
{\small{\it LMV - Laboratoire de Mathématiques de Versailles UMR8100\\
 UVSQ - Université de Versailles et Saint Quentin\\
           45, Avenue des Etats-Unis, 78035 Versailles - France}\\
         \href{mailto:paolo.vannucci@uvsq.fr}{paolo.vannucci@uvsq.fr}}}

\begin{document}
\maketitle

\hrule
\begin{abstract}
This paper focuses on the conditions for obtaining auxetic, i.e. with a negative Poisson' ratio, composite laminates made of specially orthotropic layers. In particular, the layers considered are of three types: $R_1-$orthotropic, i.e. square-symmetric plies, like those reinforced by balanced fabrics, $R_0-$orthotropic layers, like those that can be obtained with balanced fabrics having warp and weft forming an angle of $45^\circ$, and finally $r_0-$orthotropic layers, like common paper. All these types of orthotropy have mathematical and mechanical properties different by common orthotropy. As a consequence of this, the conditions of auxeticity  for anisotropic composite laminates made of such special plies change from the more common case of unidirectional plies. These conditions are analyzed in this paper making use of the polar formalism, a mathematical method particularly suited for the study of two-dimensional anisotropic elasticity.

\end{abstract}
\medskip
\hrule
\bigskip


\section{Introduction}
\label{sec:intro}

Auxetic materials are elastic bodies with a negative   Poisson's ratio $\nu$. This property  is predicted by the classical theory of elasticity also for isotropic bodies, for which ${-1<\nu<1/2}$, \cite{Love, sokolnikoff}. However, for such a class of materials, auxeticity is normally get thanks to some kind of subjacent microstructure in the material, giving rise to special kinematics in the continuum. Scientific literature is rather wide on this topic, see for instance 
\cite{cho2019,Almgren85,Evans91,Evans_Nature91,Lakes1991,Lakes1993,Lakes2002,Lakes2017,Milton92}. A review of  the state of the art can be found in\cite{Prawoto12} or in the more recent \cite{Shukla22}. 

Apart the use of materials with peculiar microstructures, there is another way to obtain a negative Poisson's ratio: anisotropy. In such a case, the variability of the elastic properties with the direction makes it possible to have negative Poisson's ratios, at least for some set of orientations, for classical elastic materials {\it à la Cauchy}, \cite{vannucci_libro}, i.e. without an underlying microstructure.
This is the case for some anisotropic materials, like for instance the pine wood cited by Lekhnitskii, \cite{Lekhnitskii}, but, more interesting for applications, auxeticity can be obtained in designed meta-materials, like anisotropic laminates. In such a case, the tailoring of the elastic properties can, for some constituent materials, lead to a negative Poisson's ratio, at least for some sets of directions. The auxeticity of laminates has been investigated in different works, from the pioneer studies of Herakovich, \cite{Herakovich84}, and Miki, \cite{Miki89}, to the experimental studies of Clarke at al\cite{Clarke94} and Hine et al\cite{Hine97} for balanced angle-ply laminates, or to the studies of Zhang et al\cite{Zhang98,Zhang99}, Alderson et al\cite{alderson2005},  concerning the possibility of obtaining auxetic laminates, or to the more recent studies on the maximization of the negative Poisson's ratio, \cite{Peel07,Shokrieh11}. The interested reader can find a recent state of the art on auxetic laminates in \cite{Veloso23}. 

In two recent papers, \cite{vannucci24b,vannucci24d}, this author considered the in-plane auxeticity of orthotropic laminates composed by identical  layers. In the first paper,  two types of auxetic laminates are introduced: the TAALs (Totally Auxetic Anisotropic Laminates) and the PAALs (Partially Auxetic Anisotropic Laminates). In the first case, the Poisson's ratio $\nud(\theta)$, defined in the next Section, is negative for each direction $\theta$, though not constant, while in the second one $\nud(\theta)<0$ only for some $\theta$. The conditions for obtaining these two types of auxetic laminates are discussed in relation to the elastic properties of the constituent ply. In the second paper, the maximization of the negative Poisson's ratio or of the set of directions with $\nud(\theta)<0$ is studied. In both these studies, like in all the literature on the subject, the constituent plies are unidirectional layers, i.e.  layers reinforced by rectilinear fibers aligned in one direction. Mechanically speaking, these plies are planar {\it ordinarily orthotropic} materials. Ordinary orthotropy  (this concept is clarified in the next Section) is a common case of orthotropy, characterized by some mechanical properties and determined by precise mathematical conditions. This affects all the mechanical properties of the laminate and of course also its auxeticity. 

However, other types of planar orthotropic materials exist, named {\it specially orthotropic}. In particular, it has been shown that three different types of specially orthotropic planar materials {\it à la Cauchy} exist: $R_1-$, $R_0-$ and $r_0-$ orthotropic planar materials. Though in all of these three cases the macroscopic mechanical properties are those typical of an orthotropic ply, actually they have some peculiarities and specific mathematical properties that render all of them {\it special}. The consequences of special orthotropy on the auxeticity of laminates is investigated in this paper. The aim of the study is to analyze how the three different types of orthotropy affect the conditions for obtaining TAALs and PAALs. In this sense, this research is purely theoretical, it aims at exploring some aspects of classical anisotropic elasticity which are still obscure. The results, hopefully, will be of some help for possible practical applications.

The study is done using a special mathematical tool, the so-called {\it polar formalism} introduced by G. Verchery, \cite{Verchery79,Meccanica05,vannucci_libro}. Such method makes use of angles and tensor invariants to describe a planar tensor of the type of elasticity, which is particularly interesting in anisotropy. In the case of specially orthotropic materials, and of their influence on auxeticity, the use of the polar formalism is practically mandatory, because it is exactly through the polar formalism that the mathematical  peculiarities of special orthotropies can be effectively represented. It is worth noting, besides, that it is exactly thanks to the polar formalism that $R_0-$ and its dual $r_0-$ orthotropies have been discovered, \cite{vannucci02joe,vannucci10joe}. 

Like in all of the papers on auxetic laminates, only the auxeticity of the in-plane behavior of extension-bending uncoupled laminates, \cite{jones,gay14,vannucci_libro}, is considered. 
This assumption is necessary:  the Poisson's ratio is  practically impossible to be analyzed  for coupled laminates, because in this case the compliance depends in a very complicate manner upon $\A,\B$ and $\D$, respectively the stiffness tensors in extension, coupling and bending\cite{vannucci01joe,vannucci23a,vannucci23b}. However, uncoupling ($\B=\mathbb{O}$) can be obtained by suitable stacking sequences,  not necessarily symmetric, \cite{vannucci01ijss,vannucci01cst}. Finally, uncoupling can be obtained rather easily and to assume it does not constitute a true limitation. Bending auxeticity is less interesting for applications; physically, it implies that the normal curvature of a plate bent along a direction is of the  the same sign of the normal curvature measured  in the orthogonal direction, i.e. that locally the curved surface of the bent plate is made of elliptic points, \cite{toponogov,vannucci_alg}.

The paper is organized as follows: in the next Section, the problem is stated using the polar formalism. In the following three Sections, the three cases of special orthotropies $R_1,R_0$ and $r_0$  are respectively treated, then some final remarks are given along with a conclusion.

\section{Mathematical formulation}
\subsection{The auxeticity condition for plane elasticity}
In a  plane elastic state, the in-plane Poisson's ratio $\nud(\theta)$  is defined as, see e.g. \cite{Lekhnitskii,TsaiHahn,jones,ting} 
\be
\nud(\theta):=-\frac{\Sy_{12}(\theta)}{\Sy_{11}(\theta)},
\ee
where $\Sy$ is the compliance   tensor and  $\theta$ is the angle measured from the $x_1-$axis. In this paper, the Kelvin's notation, \cite{kelvin,kelvin1}, for the elasticity tensors is adopted.
Because $\Sy_{11}(\theta)>0\ \forall\theta$, \cite{vannucci_libro,vannucci24a}, 
\be
\nud(\theta)<0\iff\Sy_{12}(\theta)>0.
\ee
This is the auxeticity condition for a plane elastic state.

\subsection{Auxeticity by the polar formalism}

The polar formalism is a mathematical technique to represent tensor,  in a two-dimensional space, by invariants and angles. By it, the above condition reads like
\be
\label{eq:auxeticity0}
\Sy_{12}(\theta)=-t_0+2t_1- r_0\cos4(\phi_0-\theta)>0,
\ee
where $t_0, t_1,r_0$ are non-negative tensor invariants of $\Sy$ and $\phi_0$ a polar angle, determined by the frame's choice. Because the homogenization laws for laminates apply to the reduced stiffness tensor $\Q=\Sy^{-1}$ of the layers, \cite{Lekhnitskii,ting,jones}, it is necessary to express the previous condition using the polar parameters of $\Q$:
\be
\label{eq:componentipolarisouplesse}
\begin{array}{l}
t_0=2\dfrac{T_0T_1-R_1^2}{\Delta},\ \ \ 
t_1=\dfrac{T_0^2-R_0^2}{2\Delta}, \ \ 
r_0 \mathrm{e}^{4\mathrm{i}\phi_0}=\dfrac{2}{\Delta}(R_1^2\mathrm{e}^{4\mathrm{i}\Phi_1}-T_1R_0\mathrm{e}^{4\mathrm{i}\Phi_0}),
\end{array}
\ee
with $T_0,T_1,R_0,R_1$ 
 the non negative polar invariants moduli of $\Q$ and $\Phi_0,\Phi_1$ the two polar angles of $\Q$, whose difference is the fifth invariant; the invariant function of invariants $\Delta$ is  given by
\be
\Delta=4T_1(T_0^2-R_0^2)-8R_1^2\left[T_0-R_0\cos4(\Phi_0-\Phi_1)\right].
\ee
Some simple passages give
\be
\begin{split}
r_0\cos4(\phi_0-\theta)&=\frac{2}{\Delta}[R_1^2\cos4(\Phi_1-\theta)-T_1R_0\cos4(\Phi_0-\theta)].
\end{split}
\ee
Actually, $\Delta=\det\Q>0$, so it   does not change the sign of $\Sy_{12}$ and can be ignored in the following. Finally, in terms of the polar parameters of $\Q$, 
the auxeticity condition (\ref{eq:auxeticity0}) is
\be
\label{eq:auxeticity1}
2(T_0T_1-R_1^2)-T_0^2+R_0^2+2[R_1^2\cos4(\Phi_1-\theta)-T_1R_0\cos4(\Phi_0-\theta)]<0.
\ee
A symmetry of the plane elastic  behavior exists if and only if the following polar condition is satisfied, \cite{vannucci_libro}:
\be
R_0R_1^2\sin4(\Phi_0-\Phi_1)=0.
\ee
As it can be seen, the existence of a form of elastic symmetry depends upon special values taken by the invariants $R_0, R_1$ or $\Phi_0-\Phi_1$. In particular, the condition 
\be
\label{eq:orth1}
\Phi_0-\Phi_1=K\frac{\pi}{4},\ \ K\in\{0,1\},
\ee
determines the {\it ordinary orthotropy}, that can hence be of two types, according to the value of $K$, $0$ or $1$.  The conditions
\be
\label{eq:orth2}
R_0=0,\ \ \ \ R_1=0
\ee
correspond to two {\it special orthotropies},  respectively $R_0-${\it orthotropy} and $R_1-${\it orthotropy}.  The last one is well known: it corresponds to the so-called {\it square symmetry}, i.e. to a case where all the elastic properties are periodic of $\pi/2$ and there are two couples of mutually orthogonal symmetry axes rotated of $\pi/4$. Unlike this one, $R_0$-orthotropy was unknown before the use of the polar formalism, \cite{vannucci02joe}. Because $R_0-$orthotropy does not imply the same for the corresponding polar parameter in compliance, $r_0$, i.e. $R_0=0\nLeftrightarrow r_0=0$, then it exists a third case of special orthotropy, \cite{vannucci10joe}, that is indicated here as $r_0-${\it orthotropy}, corresponding to the condition
\be
\label{eq:orth3}
r_0=0.
\ee
This does not happen for square symmetry, because $R_1=0\Leftrightarrow r_1=0$, with $r_1$ the compliance polar corresponding of $R_1$. 

It is worth considering the differences between ordinary orthotropy and the three cases of special orthotropies. Algebraically speaking, condition (\ref{eq:orth1}) involves a cubic invariant, while (\ref{eq:orth2}) and (\ref{eq:orth3}) concern quadratic invariants. Moreover, it is important to recall that the polar formalism decomposes elasticity as a sum of harmonics, \cite{vannucci_libro}: the isotropic harmonic, independent from the orientation $\theta$, represented, for the stiffness tensor, by the invariants $T_0$ and $T_1$, the harmonic varying with circular functions of  $4\theta$, represented by the invariant $R_0$ and the angle $\Phi_0$, and the harmonic varying with $2\theta$, represented by $R_1$ and the angle $\Phi_1$. Analogous considerations can be done for the compliance (whose polar parameters are similar but conventionally denoted by lower-case letters).  

Ordinary orthotropy corresponds hence to a particular value of the difference $\Phi_0-\Phi_1$  of the two polar angles: when this difference, that is a tensor invariant, is either $0$ or $\pi/4$, i.e. when the two harmonics varying like $4\theta$ and $2\theta$ are in phase or shifted of $\pi/4$, then the behavior is ordinarily orthotropic. However orthotropy can exists also when one of the two harmonics disappear, i.e. when either $R_0=0$ or $R_1=0$. In these cases, the angular variation of the elastic moduli is no longer like in the case of ordinary orthotropy: as already mentioned, when $R_1=0$, all the moduli vary like $4\theta$ and are periodic of $\pi/2$. When $R_0=0$, all the moduli vary like $2\theta$, i.e. like the components of a second rank tensor and some of them, namely $\Q_{12}$ and $\Q_{66}$, are isotropic. When this is the case for compliance, i.e. when $r_0=0$, then $\Sy_{12}$ and $\Sy_{66}$ are isotropic, the last giving also the isotropy of the shear modulus $G_{12}$. This is the case of a rather common material: paper, \cite{vannucci10joe}. 

Finally, specially orthotropic materials have mathematical and mechanical properties rather different from the ordinarily orthotropic ones, which affects the design of laminates, their final properties. It is to be expected, hence, that special orthotropies affect also the design of auxetic laminates. This is the topic of the study developed in the following Sections, in the order for laminates composed of identical $R_1-$, $R_0-$ or $r_0-$orthotropic layers.

To this purpose, it is worth to recall the homogenization laws giving the components of the in-plane elastic stiffness tensor $\A$ for a laminate composed by identical layers, that for  the polar formalism are, \cite{vannucci_libro},
\be
\begin{array}{l}
\label{eq:homoga}
T_0^\A=T_0,\medskip\\
T_1^\A=T_1,\medskip\\
R_0^\A\ \e^{4\i\Phi_0^\A}=R_0\ \e^{4\i\Phi_0}(\xi_1+\i \xi_2),\medskip\\
R_1^\A\ \e^{2\i\Phi_1^\A}=R_1\ \e^{2\i\Phi_1}(\xi_3+\i \xi_4).
\end{array}
\ee
A polar component for the laminate is indicated by an apex $\A$, while the polar parameters of the layer do not have any apex. The quantities $\xi_1$ to $\xi_4$, the so-called {\it lamination parameters}, \cite{tsai1968}, are given, for a laminate of $n$ plies with orientation angles $\delta_j$, by
\be
\label{eq:lampar}
\xi_1+\i\xi_2=\frac{1}{n}\sum_{j=1}^n\e^{4\i\delta_j},\ \ \ \xi_3+\i\xi_4=\frac{1}{n}\sum_{j=1}^n\e^{2\i\delta_j}.
\ee 
It is worth noticing that the isotropic part of the layer and of $\A$, i.e. of the laminate, are the same, and it cannot be affected by  the orientations of the layers.  This is a consequence of the fact that layers are identical. Also, it is apparent that 
\be
R_0=0\Rightarrow R_0^\A=0,\ \ \ \ R_1=0\Rightarrow R_1^\A=0,
\ee
i.e. the special symmetries $R_0=0$ and $R_1=0$ are preserved by laminate's homogenization, which is not the case for ordinary orthotropy nor for $r_0=0$ special orthotropy (in this last case because homogenization laws like (\ref{eq:homoga}) apply only to stiffness properties, not to the compliance ones). 

Finally, the auxeticity condition for having a negative Poisson's ratio $\nud^\A$ for the extension behavior of the laminate, is 
\be
\label{eq:aux1}
\bs
2(T_0T_1-{R_1^\A}^2)-T_0^2+{R_0^\A}^2+2\left[{R_1^\A}^2\cos4(\Phi_1^\A-\theta)-T_1R_0^\A\cos4(\Phi_0^\A-\theta)\right]<0.
\end{split}
\ee

\section{Laminates composed of \bf{$R_1-$}orthotropic layers}
\subsection{Dimensionless auxeticity condition}
As first case, laminates composed by square-symmetric identical layers are considered: $R_1=0\Rightarrow R_1^\A=0$. 
Moreover, multiplying eq. (\ref{eq:homoga})$_3$ by $\e^{-4\i\Phi_0}$ we get
\be
R_0^\A\e^{4\i(\Phi_0^\A-\Phi_0)}=R_0(\xi_1+\i\xi_2).
\ee
If the choice 
\be
\label{eq:angles0}
\Phi_0^\A-\Phi_0=c_0\frac{\pi}{2},\ \ \ c_0\in\mathbb{N},
\ee
is done, which is  always possible, because $\Phi_0^\A$ and $\Phi_0$ just fixe the reference frame, and because $R_0^\A,R_0,\xi_1,\xi_2\in\mathbb{R},$ then 
\be
\label{eq:R0laminato}
\xi_2=0\ \Rightarrow\ R_0^\A=R_0\ \xi_1,
\ee
with
\be
\xi_1=\frac{1}{n}\sum_{j=1}^n\cos4\delta_j,\ \ \ \xi_1\in[-1,1].
\ee
Indeed, the above condition (\ref{eq:angles0}) simply fixes a  reference frame for the laminate wherein it is particularly simple to write the equations. Finally, if a rotation of $\Phi_0^\A$ is done, the auxeticity condition for the in-plane behavior of the laminate becomes
\be
\label{eq:aux2}
2T_0T_1-T_0^2+R_0^2\ \xi_1^2-2T_1R_0\ \xi_1\cos4\theta<0.
\ee
In order to simplify the analytical developments, we observe that adding or subtracting the angle $\pi/4$ to the orientation $\delta_j$ of each ply (i.e. rotating the frame of $\pm\pi/4$) simply changes the sign of $\xi_1$, so it is sufficient to bound the analysis to the set $\xi_1\in[0,1]$. Then, we introduce the dimensionless quantities
\be
\label{eq:paradim}
\tau:=\frac{T_1}{T_0},\ \ \ \rho:=\frac{R_0}{T_0}.
\ee
Because, see \cite{vannucci_libro}, the elastic bounds on the polar parameters for the case $R_1=0$ are
\be
\label{eq:elboundpol}
T_1>0,\ \ \ 0<R_0<T_0,
\ee
then the bounds defining the elastic domain in the plane $(\tau,\rho)$ are
\be
\label{eq:elboundpoladim}
\tau>0,\ \ \ 0<\rho<1.
\ee
Finally, eq. (\ref{eq:aux2}) becomes
\be
\label{eq:aux3}
\mu(\xi_1,\theta):=2\tau-1+\rho^2\xi_1^2-2\tau \rho\ \xi_1\cos4\theta<0.
\ee

\subsection{Totally Auxetic Anisotropic Laminates (TAALs)}
\subsubsection{Polar conditions}
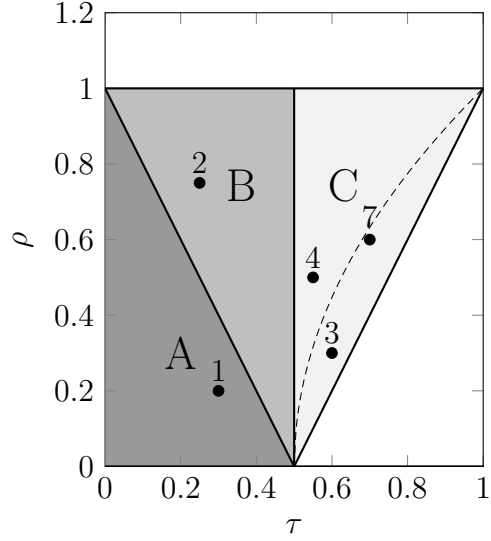
\begin{figure}
\begin{center}
\begin{tikzpicture}
\begin{axis} [xmin=0,xmax=1.,xlabel=$\tau$,
 ymin=0,ymax=1.2,ylabel=$\rho$, width=5cm,height=6cm,scale only axis]
 \addplot[ thick,name path=A,domain=0:0.5] {1-2*x};
 \addplot[thick, name path=B] {0};
\addplot[domain=0:1,  thick,name path=C]coordinates{(0,1)(0.5,1)};
\addplot[ thick,name path=D]coordinates{(0.5,1)(1,1)};
\addplot[ thick]coordinates{(0.5,0)(0.5,1)};
\addplot[ thick,name path=E,domain=0.5:1]{2*x-1};
\addplot [only marks,domain=0:1] coordinates{(0.3,0.2)(0.25,0.75)(0.6,0.3)(0.55,0.5)(0.7,0.6)};
\addplot[gray,fill opacity=0.8] fill between[of=A  and B] ;
\addplot[gray,fill opacity=0.5] fill between[of=A  and C] ;
\addplot[gray,fill opacity=0.1] fill between[of=E  and D] ;
\pgfplotsset{every axis/.append style={extra description/.code={\node at (0.2,0.25) {\Large{A}};\node at (0.36,0.62){\Large{B}};\node at (0.63,0.62){\Large{C}};\node at (0.3,0.21){$1$};\node at (0.25,0.67){$2$};\node at (0.6,0.295){$3$};\node at (0.55,0.46){$4$};\node at (0.7,0.55){$7$};}}}
\addplot[domain=0.5:1,samples=100,densely dashed]{(2*x-1)^0.5};
\end{axis}
\end{tikzpicture}
\caption{Domains A and B  of solutions for TAALs and domain C of solutions for PAALs in the plane $(\tau,\rho)$; the dotted line is the graphic of curve $\rho=\sqrt{2\tau-1}$.}
\label{fig:1}
\end{center}
\end{figure}

To obtain a TAAL, eq. (\ref{eq:aux3}) must be satisfied $\forall\theta$. Because $\mu(\xi_1,\theta)$ gets its minimum  for $\theta=\pi/4$, the condition for having a TAAL is
\be
\hat{\mu}(\xi_1)=2\tau-1+\rho^2\xi_1^2+2\tau \rho\ \xi_1<0,\ \ \ \xi_1\in[0,1].
\ee
The possible solutions to the above inequality give the admissibles sets for $\xi_1$. Usually, a value of $\xi_1$ can be obtained by different stacking sequences of layers. So, if the set of admissible $\xi_1$s is not empty, TAALs composed of layers with given values  of $\tau$ and $\rho$ can be fabricated. 
The bounds of the admissible set given by the above inequality are
\be
(\xi_1)_{1,2}=\frac{-\tau\pm|\tau-1|}{\rho}.
\ee
Considering that 
$\xi_1\in[0,1]$ and looking for solutions in the elastic domain  defined by the bounds in eq. (\ref{eq:elboundpoladim}), to be satisfied by any material, some simple though articulated steps give finally that TAALs are possible:
\begin{enumerate}[a)]
\item $\forall\xi_1\in[0,1]$   if  the couple  $(\tau,\rho)$ belongs to the subset  A of points satisfying the condition
\be
\label{eq:setA}
\tau<\frac{1-\rho}{2};
\ee
\item for $\xi_1\in\left[0,\dfrac{1-2\tau}{\rho}\right]$ if the couple $(\tau,\rho)$ belongs to the subset  B of points satisfying the conditions
\be
\label{eq:setB}
\frac{1-\rho}{2}<\tau<\frac{1}{2}
\ee
\end{enumerate}
see Fig. \ref{fig:1}. It is worth to remark that the only physical bounds on $\tau$ and $\rho$ are those in eq. (\ref{eq:paradim}), hence all the points in the sets A and B are physically admissible.


Conditions (\ref{eq:setA}) and (\ref{eq:setB}) can be expressed using  the polar components $T_0,T_1,R_0$, to obtain 
\be
\begin{array}{l}
\label{eq:setAB}
\mathrm{-\ for\ set\  A:}\ \ \ T_1<\dfrac{T_0-R_0}{2}
;\medskip\\
\mathrm{-\ for\  set\  B:}\ \ \ \dfrac{T_0-R_0}{2}<T_1<\dfrac{T_0}{2},
\end{array}
\ee
the elastic domain now being defined by bounds (\ref{eq:elboundpol}). Together with eqs. (\ref{eq:setA}) and (\ref{eq:setB}), that are in dimensionless form, these bounds define in the polar formalism the conditions that a ply must satisfy to fabricate a TAAL.

\subsubsection{Cartesian conditions}
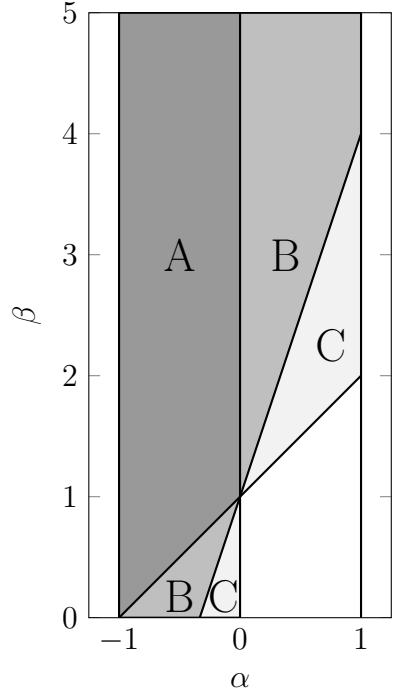
\begin{figure}
\begin{center}
\begin{tikzpicture}
\begin{axis} [xmin=-1.25,xmax=1.25,xlabel=$\alpha$,
 ymin=0,ymax=5,ylabel=$\beta$, width=4cm,height=8cm,scale only axis]
 \addplot[ thick]coordinates{(-1,0)(-1,5)} ;
 \addplot[ thick,name path=1a,domain=-1:0]{1+x};
  \addplot[ thick,name path=1b,domain=0:1]{1+x};
 \addplot[ thick,name path=2a,domain=-1/3:0]{1+3*x};
\addplot[ thick,name path=2b,domain=0:1]{1+3*x};
\addplot[ thick,name path=3a,domain=-1:-1/3]{0};
\addplot[ thick,name path=3b,domain=-1/3:0]{0};
\addplot[ thick,name path=4a,domain=-1:0]{5};
\addplot[ thick,name path=4b,domain=0:1]{5};
 \addplot[ thick]coordinates{(0,0)(0,5)} ;
 \addplot[ thick]coordinates{(1,0)(1,5)} ;
 \addplot[gray,fill opacity=0.8] fill between[of=1a  and 4a] ;
\addplot[gray,fill opacity=0.5] fill between[of=1a  and 2a] ;
\addplot[gray,fill opacity=0.5] fill between[of=4b  and 2b] ;
\addplot[gray,fill opacity=0.1] fill between[of=3b  and 2a] ;
\addplot[gray,fill opacity=0.1] fill between[of=1b  and 2b] ;
\pgfplotsset{every axis/.append style={extra description/.code={\node at (0.3,0.6) {\Large{A}};\node at (0.65,0.6){\Large{B}};\node at (0.8,0.45){\Large{C}};\node at (0.3,0.035){\Large{B}};\node at (0.445,0.035){\Large{C}};}}}
\end{axis}
\end{tikzpicture}
\caption{Domains A and B  of solutions for TAALs and domain C of solutions for PAALs in the plane $(\alpha,\beta)$.}
\label{fig:2}
\end{center}
\end{figure}

The above polar condition can be traduced into conditions using the Cartesian components of the reduced stiffness tensor $\Q$. To this end, the relation between the polar parameters and the components of $\Q$ for an orthotropic ply are, \cite{vannucci_libro}, (the Kelvin's notation is used for the components $\Q_{ij}$, \cite{kelvin,kelvin1}):
\be
\label{eq:polcart}
\begin{array}{l}
T_0=\dfrac{1}{8}(\Q_{11}-2\Q_{12}+2\Q_{66}+\Q_{22}),\medskip\\
T_1=\dfrac{1}{8}(\Q_{11}+2\Q_{12}+\Q_{22}),\medskip\\
R_0=\dfrac{1}{8}|\Q_{11}-2\Q_{12}-2\Q_{66}+\Q_{22}|,\medskip\\
R_1=\dfrac{1}{8}|\Q_{11}-\Q_{22}|,\medskip\\
\Phi_0=0\ \mathrm{if}\ \Q_{11}-2\Q_{12}-2\Q_{66}+\Q_{22}>0\ \mathrm{otherwise}\ \Phi_0=\dfrac{\pi}{4},
\medskip\\
\Phi_1=0 \ \mathrm{if}\ \Q_{11}-\Q_{22}>0\ \mathrm{otherwise}\ \Phi_1=\dfrac{\pi}{2}.
\end{array}
\ee
So, replacing these relations into eq. (\ref{eq:setAB}) and considering that $R_1=0\Rightarrow\Q_{11}=\Q_{22}$, gives
\be
\label{eq:condsetAB}
\begin{array}{l}
- \mathrm{for\ set\ A:}\ \ \ \Q_{66}>\Q_{11}+3\Q_{12}+|\Q_{11}-\Q_{12}-\Q_{66}|<\Q_{66},\medskip\\
- \mathrm{for\ set\ B:}\ \ \ \Q_{11}+3\Q_{12}<\Q_{66}<\Q_{11}+3\Q_{12}+ |\Q_{11}-\Q_{12}-\Q_{66}|.
\end{array}
\ee
 Also in this case it is worth to pass to dimensionless moduli. To this end, remembering that $\Q_{11}>0$, the following parameters are introduced:
\be
\label{eq:alphabeta}
\alpha:=\frac{\Q_{12}}{\Q_{22}}=\frac{\Q_{12}}{\Q_{11}},\ \ \ \beta:=\frac{\Q_{66}}{\Q_{22}}=\frac{\Q_{66}}{\Q_{11}}.
\ee
Because, cf. \cite{vannucci_libro}, for materials with $R_1=0$
\be
\Q_{12}=\nud\Q_{11}=\nud\Q_{22},
\ee
and, see \cite{vannucci23c},
\be
\Q_{66}>0,\ \ \ -1<\nud<1,
\ee
it is (to remark that actually $\alpha=\nud$)
\be
-1<\alpha<1,\ \ \ \beta>0.
\ee
Then, bounds (\ref{eq:condsetAB}) can be rewritten in dimensionless form:
\be
\label{eq:setABalphabeta}
\begin{array}{l}
- \mathrm{for\ set\ A:}\ \ \ \beta>1+3\alpha+|1-\alpha-\beta|,\medskip\\
- \mathrm{for\ set\ B:}\ \ \ 1+3\alpha<\beta<1+3\alpha+|1-\alpha-\beta|.
\end{array}
\ee
In Fig. \ref{fig:2} the domains A and B are represented in the space $(\alpha,\beta)$. Because only materials with $\alpha<0\Rightarrow\nud<0\Rightarrow\Q_{12}<0$ belong to set A, it is possible to have a TAAL $\forall\xi_1$ only using a ply that is totally auxetic by itself. In fact, a special case is a laminate obtained superposing $n$ layers all with the same orientation. In such a case $\A(\theta)=\Q(\theta)$, so to have $\nud^\A(\theta)<0\ \forall\theta$ it must be $\nud(\theta)<0\ \forall\theta$.


\subsubsection{Technical moduli conditions}
A further step consists in expressing the auxeticity conditions  through the technical constants. For an orthotropic layer it is
\be
\label{eq:Qorthply}
\begin{array}{l}
\Q_{11}=\dfrac{E_1}{1-\nu_{12}\nu_{21}},\medskip\\
\Q_{12}=\dfrac{\nu_{12}E_2}{1-\nu_{12}\nu_{21}},\medskip\\
\Q_{22}=\dfrac{E_2}{1-\nu_{12}\nu_{21}},\medskip\\
\Q_{66}=2G_{12}.
\end{array}
\ee
By the reciprocity condition
\be
\label{eq:reciprocity}
\frac{\nu_{12}}{E_1}=\frac{\nu_{21}}{E_2},
\ee
because $R_1=0\Rightarrow\Q_{11}=\Q_{22}$, it is $E_1=E_2,\nu_{12}=\nu_{21}$ and then
\be
\label{eq:Qijs}
\Q_{11}=\Q_{22}=\dfrac{E_1}{1-\nu_{12}^2},\ \ \Q_{12}=\dfrac{\nud E_1}{1-\nu_{12}^2},\ \ \Q_{66}=2G_{12}.
\ee
Finally, eq. (\ref{eq:condsetAB}) becomes
\be
\label{eq:condEiGij}
\begin{array}{l}
- \mathrm{for\ set\ A:}\ \ \ G_{12}>\dfrac{1+3\nud}{2(1-\nu_{12}^2)}E_1+\left|\dfrac{E_1}{2(1+\nud)}-G_{12}\right|,\medskip\\
- \mathrm{for\ set\ B:}\ \ \ \dfrac{1+3\nud}{2(1-\nu_{12}^2)}E_1<G_{12}<\dfrac{1+3\nud}{2(1-\nu_{12}^2)}E_1+\left|\dfrac{E_1}{2(1+\nud)}-G_{12}\right|.
\end{array}
\ee
\begin{figure}
\begin{center}
\begin{tikzpicture}
\begin{axis} [xmin=-1.25,xmax=1.25,xlabel=$\alpha$,
 ymin=0,ymax=2.5,ylabel=$\gamma$, width=5cm,height=5cm,scale only axis]
\addplot[ thick] coordinates{(-1,0)(-1,5)} ;
\addplot[ thick]coordinates{(0,0)(0,5)} ;
\addplot[ thick,name path=1a,domain=-1:0]{1/(2*(1-x))};
\addplot[ thick,name path=1b,domain=0:0.83]{1/(2*(1-x))};
\addplot[ thick,name path=2a,domain=-1:-1/3]{0};
\addplot[ thick,name path=2b,domain=-1/3:0]{(1+3*x)/(2*(1-x^2))};
\addplot[ thick,name path=2c,domain=-1/3:0]{0};
\addplot[ thick,name path=3,domain=0:0.83]{(1+3*x)/(2*(1-x^2))};
\addplot[ thick,name path=4a,domain=-1:0]{5};
\addplot[ thick,name path=4b,domain=0:1]{5};
\addplot[gray,fill opacity=0.8] fill between[of=1a  and 4a] ;
\addplot[gray,fill opacity=0.5] fill between[of=1a  and 2a] ;
\addplot[gray,fill opacity=0.5] fill between[of=4b  and 3] ;
\addplot[gray,fill opacity=0.1] fill between[of=1b and 3];
\addplot[gray,fill opacity=0.1] fill between[of=2b and 2c];
\pgfplotsset{every axis/.append style={extra description/.code={\node at (0.3,0.6) {\Large{A}};\node at (0.6,0.6){\Large{B}};\node at (0.68,0.47){\Large{C}};\node at (0.3,0.055){\Large{B}};\node at (0.454,0.055){\Large{C}};}}}
\end{axis}
\end{tikzpicture}
\caption{Domains A and B  of solutions for TAALs and domain C of solutions for PAALs in the plane $(\alpha,\gamma)$.}
\label{fig:3}
\end{center}
\end{figure}
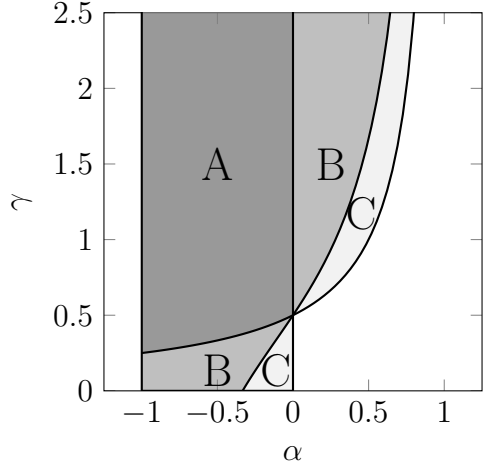
Also in this case, it is worth to have a dimensionless representation of the domains A and B. To this purpose, let us introduce the dimensionless parameter
\be
\label{eq:gammadef}
\gamma:=\frac{G_{12}}{E_2}=\frac{G_{12}}{E_1},\ \ \gamma>0.
\ee
Then, the above bounds can be rewritten as
\be
\begin{array}{l}
\label{eq:boundsgammaalpha}
- \mathrm{for\ set\ A:}\ \ \ \gamma>\dfrac{1+3\alpha}{2(1-\alpha^2)}+\left|\dfrac{1}{2(1+\alpha)}-\gamma\right|,\medskip\\
- \mathrm{for\ set\ B:}\ \ \ \dfrac{1+3\alpha}{2(1-\alpha^2)}<\gamma<\dfrac{1+3\alpha}{2(1-\alpha^2)}+\left|\dfrac{1}{2(1+\alpha)}-\gamma\right|.
\end{array}
\ee
The domains A and B in the space $(\alpha,\gamma)$ are shown in Fig. \ref{fig:3}.
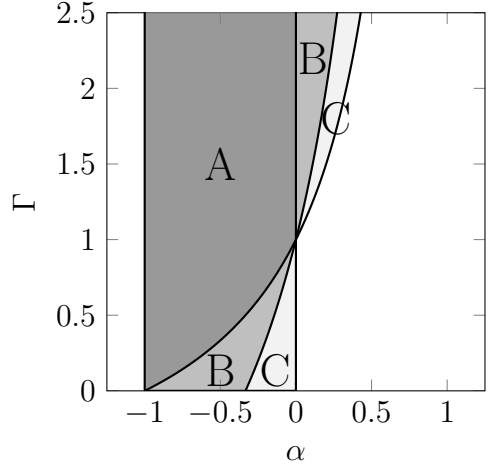
\begin{figure}
\begin{center}
\begin{tikzpicture}
\begin{axis} [xmin=-1.25,xmax=1.25,xlabel=$\alpha$,
 ymin=0,ymax=2.5,ylabel=$\Gamma$, width=5cm,height=5cm,scale only axis]
\addplot[ thick] coordinates{(-1,0)(-1,5)} ;
\addplot[ thick]coordinates{(0,0)(0,5)} ;
\addplot[ thick,name path=1a,domain=-1:0]{(1+x)/(1-x)};
\addplot[ thick,name path=1b,domain=0:0.5]{(1+x)/(1-x)};
\addplot[ thick,name path=2a,domain=-1:-1/3]{0};
\addplot[ thick,name path=2b,domain=-1/3:0]{(1+3*x)/(1-x)};
\addplot[ thick,name path=2c,domain=-1/3:0]{0};
\addplot[ thick,name path=3,domain=0:0.34]{(1+3*x)/(1-x)};
\addplot[ thick,name path=4a,domain=-1:0]{5};
\addplot[ thick,name path=4b,domain=0:1]{5};
\addplot[gray,fill opacity=0.8] fill between[of=1a  and 4a] ;
\addplot[gray,fill opacity=0.5] fill between[of=1a  and 2b] ;
\addplot[gray,fill opacity=0.5] fill between[of=4b  and 3] ;
\addplot[gray,fill opacity=0.1] fill between[of=1b  and 3] ;
\addplot[gray,fill opacity=0.1] fill between[of=2c  and 2b] ;
\pgfplotsset{every axis/.append style={extra description/.code={\node at (0.3,0.6) {\Large{A}};\node at (0.545,0.88){\Large{B}};\node at (0.605,0.72){\Large{C}};\node at (0.3,0.055){\Large{B}};\node at (0.445,0.055){\Large{C}};}}}
\end{axis}
\end{tikzpicture}
\caption{Domains A and B  of solutions for TAALs and domain C of solutions for PAALs in the plane $(\alpha,\Gamma)$.}
\label{fig:4}
\end{center}
\end{figure}

\subsubsection{Further considerations and examples}
It is apparent that the  simplest way to define sets A and B is to use the polar formalism. 
Also, it is interesting to remark that 
\be
G_{iso}:=\dfrac{E_1}{2(1+\nu_{12})}
\ee
can be interpreted as the shear modulus of an isotropic material with $E_1$ and $\nu_{12}$ as Young's modulus and Poisson's ratio, respectively. But, see \cite{vannucci_libro}, 
\be
T_0=\dfrac{1}{8}\left(\dfrac{E_1(E_1+E_2-2\nu_{12}E_2)}{E_1-\nu_{12}^2E_2}+4G_{12}\right),
\ee
and because $R_1=0\rightarrow E_1=E_2$, then
\be
T_0=\frac{G_{12}+G_{iso}}{2},
\ee 
i.e. $T_0$ is the mean between $G_{12}$ and $G_{iso}$. If now the dimensionless parameter 
\be
\Gamma:=\frac{G_{12}}{G_{iso}}
\ee
is introduced, then the bounds (\ref{eq:condEiGij}) can be rewritten also as
\be
\label{boundsalphaGamma}
\begin{array}{l}
- \mathrm{for\ set\ A:}\ \ \ \Gamma>\dfrac{1+3\alpha}{1-\alpha}+\left|1-\Gamma\right|,\medskip\\
- \mathrm{for\ set\ B:}\ \ \  \dfrac{1+3\alpha}{1-\alpha}<\Gamma<\dfrac{1+3\alpha}{1-\alpha}+\left|1-\Gamma\right|.
\end{array}
\ee
The domains A and B in the space $(\alpha,\Gamma)$ are depicted in Fig. \ref{fig:4}.

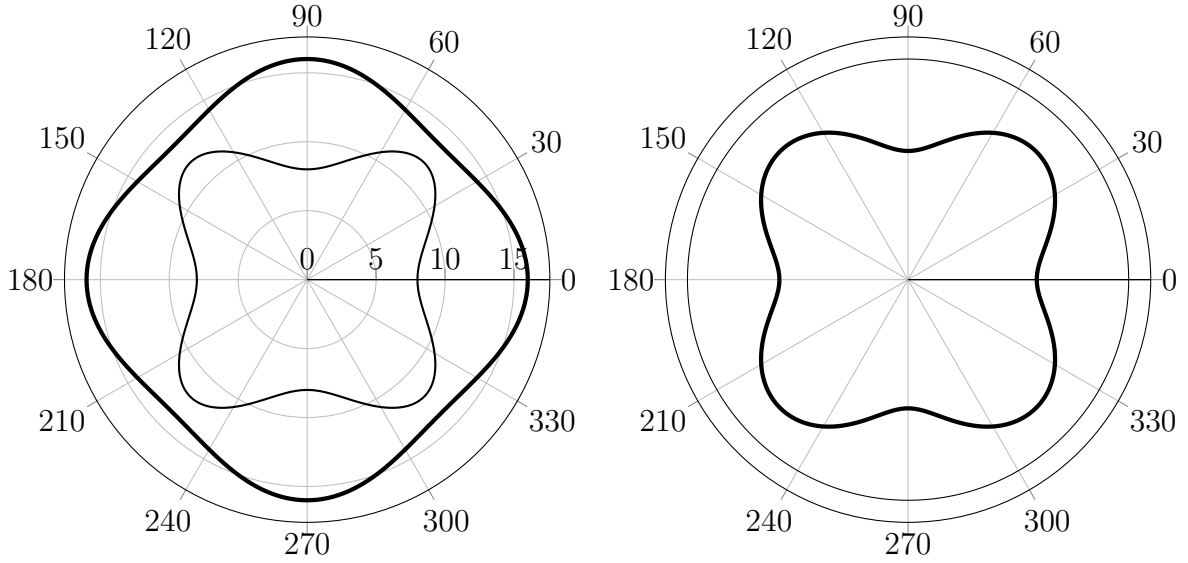
\begin{figure}[h]
\begin{center}
\begin{tikzpicture}
\begin{polaraxis} [width=0.5\textwidth]
 \addplot[ultra thick,domain=0:360,samples=360]
{0.001/(0.0000677083+0.00000520833*cos(4*(45-x)))};
\addplot[thick,domain=0:360,samples=360]
{0.001/(4*(0.0000260417-0.00000520833*cos(4*(45-x))))};
\end{polaraxis}
\end{tikzpicture}
\begin{tikzpicture}
\begin{polaraxis} [width=0.5\textwidth,ytick={-0.5}]
 \addplot[ultra thick,domain=0:360,samples=360]
{0.8+(-0.0000156249+0.00000520833*cos(4*(45-x)))/
(0.0000677083+0.00000520833*cos(4*(45-x)))};
\addplot[ domain=0:360,samples=360]
{0.8};
\end{polaraxis}
\end{tikzpicture}
\caption{Polar diagrams for the layer of Example 1; left: $E_1(\theta),G_{12}(\theta)$, respectively thick and thin curves; right:  $\nu_{12}(\theta)$ (the thin circular line corresponds to zero: inside it, $\nud<0$).}
\label{fig:5}
\end{center}
\end{figure}

Let us now consider two examples:
 
{\bf Example 1}: the first example is that of a ply whose characteristics are, in given units, $E_1=E_2=16, G_{12}=8,\nu_{12}=-0.333\Rightarrow\Q_{11}=\Q_{22}=18,\Q_{12}=-6,\Q_{66}=16$, which gives $T_0=10,T_1=3,R_0=2,\Phi_0=0\Rightarrow\tau=0.3,\rho=0.2$, so a point of set A, indicated by label "1" in Fig. \ref{fig:1}. It is easy to check that bounds (\ref{eq:setAB})$_1$ are satisfied, like also all the other equivalent bounds (\ref{eq:setABalphabeta})$_1$,  (\ref{eq:condsetAB}$_1$, (\ref{eq:condEiGij})$_1$, (\ref{eq:boundsgammaalpha})$_1$ and (\ref{boundsalphaGamma})$_1$. In Fig. \ref{fig:5} the polar diagrams of $E_1(\theta),G_{12}(\theta)$ and $\nu_{12}(\theta)$ are shown; as apparent, $\nu_{12}(\theta)<0\ \forall\theta$. A laminate realized with such a material will also have $\nud^\A(\theta)<0\ \forall\theta$, for any possible choice of $\xi_1$, i.e. for any stacking sequence. A particular, but intriguing case, is that of $\xi_1=0$, i.e. of an isotropic laminate, that can be fabricated using a symmetric stack of the Werren and Norris type, \cite{werren53}, for instance the sequence $[0^\circ,60^\circ,-60^\circ]_{sym}$. In that case, $\forall\theta$ it is $E_1^\A=15,G_{12}^\A=10,\nud^\A=-0.25$.

 {\bf Example 2}: the second example is that of a ply with  $E_1=E_2=6.667,G_{12}=17.5,\nu_{12}=0.333\Rightarrow\Q_{11}=\Q_{22}=7.5,\Q_{12}=2.5,\Q_{66}=35$, to which correspond the polar parameters $T_0=10, T_1=2.5, R_0=7.5\Rightarrow\tau=0.25,\rho=0.75$, i.e. a point of the set B, indicated by label "2" in Fig. \ref{fig:1}. The polar diagrams of the Young's modulus, of the shear modulus  and of the Poisson's ratio  for the ply are in Fig. \ref{fig:6}, while those of a laminate with $\xi_1=1/2$ are shown in Fig. \ref{fig:7}: it is apparent that $\nud^\A<0\ \forall\theta$. To remark that such a laminate can be realized, for instance, by an angle-ply sequence, i.e. by a stack with plies at the orientation $\pm\delta$, in equal number. Then, $\xi_1=\cos4\delta\Rightarrow\delta=\pi/12=15^\circ$. Also in this case, if $\xi_1=0$, i.e. with the same isotropic sequence above, we should get a laminate with $E_1^\A=13.3,G_{12}^\A=10$ and $\nud^\A=-0.33\ \forall\theta$.
\begin{figure}[h]
\begin{center}
\begin{tikzpicture}
\begin{polaraxis} [width=0.5\textwidth]
 \addplot[ultra thick,domain=0:360,samples=360]
{0.001/(0.000107+0.0000429*cos(4*x))};
\addplot[thick,domain=0:360,samples=360]
{0.001/(4*(0.00005714-0.00004286*cos(4*x))};
\end{polaraxis}
\end{tikzpicture}
\begin{tikzpicture}
\begin{polaraxis} [width=0.5\textwidth,ytick={-0.5}]
 \addplot[ultra thick,domain=0:360,samples=360]
{0.8+(0.000007143+0.0000429*cos(4*x))/(0.000107+0.0000429*cos(4*x))};
\addplot[ domain=0:360,samples=360]
{0.8};
\end{polaraxis}
\end{tikzpicture}
\caption{Polar diagrams for the layer of Example 2; left: $E_1(\theta),G_{12}(\theta)$, respectively thick and thin curves; right:  $\nu_{12}(\theta)$ (the thin circular line corresponds to zero: inside it, $\nud<0$).}
\label{fig:6}
\end{center}
\end{figure}
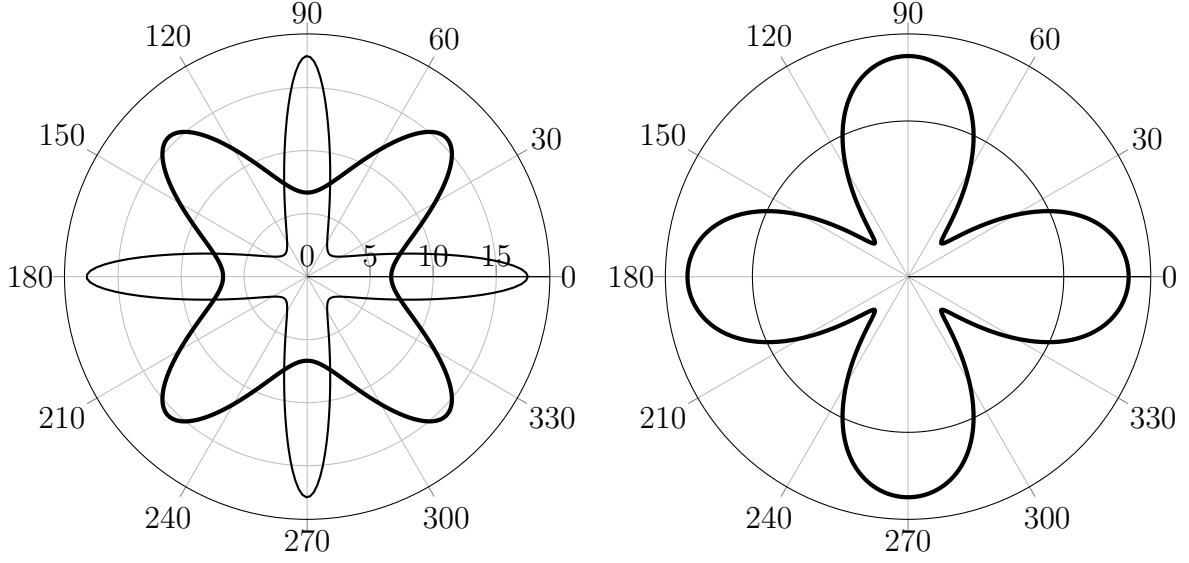
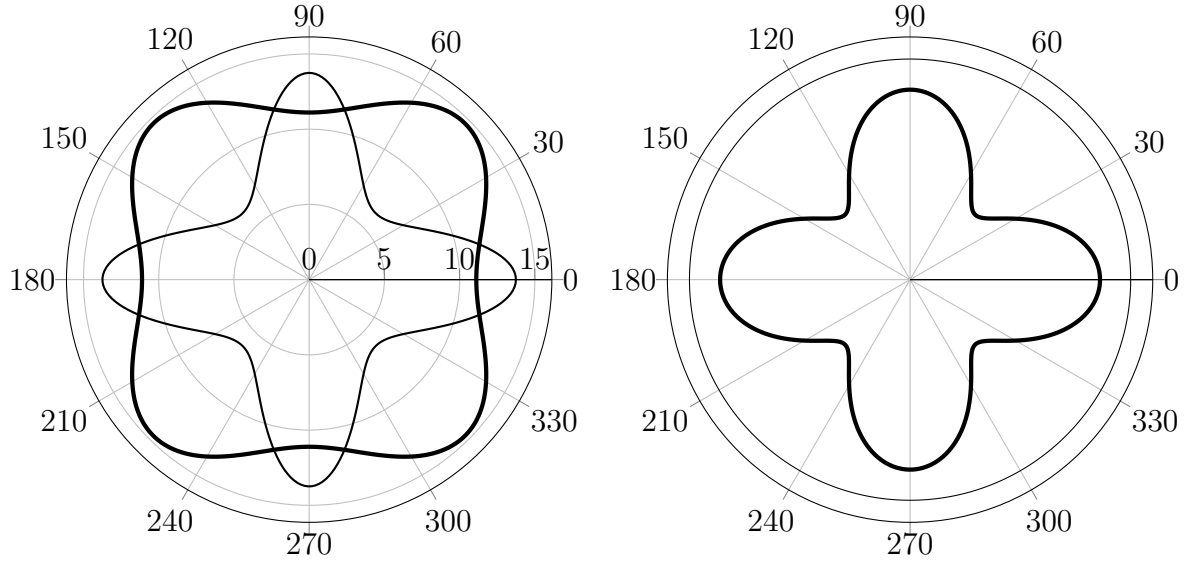
\begin{figure}
\begin{center}
\begin{tikzpicture}
\begin{polaraxis} [width=0.5\textwidth]
 \addplot[ultra thick,domain=0:360,samples=360]
{0.001/(0.0000791+0.00001091*cos(4*x))};
\addplot[thick,domain=0:360,samples=360]
{0.001/(4*(0.0000291-0.00001091*cos(4*x))};
\end{polaraxis}
\end{tikzpicture}
\begin{tikzpicture}
\begin{polaraxis} [width=0.5\textwidth,ytick={-0.5}]
 \addplot[ultra thick,domain=0:360,samples=360]
{0.8+(-0.00002091+0.00001091*cos(4*x))/(0.0000791+0.00001091*cos(4*x))};
\addplot[ domain=0:360,samples=360]
{0.8};
\end{polaraxis}
\end{tikzpicture}
\caption{{Polar diagrams for the laminate with $\xi_1=1/2$ made with the material of Example 2; left: $E_1^\A(\theta),G_{12}^\A(\theta)$, respectively thick and thin curves; right:  $\nu_{12}^\A(\theta)$ (the thin circular line corresponds to zero: inside it, $\nud^\A<0$).}}
\label{fig:7}
\end{center}
\end{figure}

The above results show that it is physically admissible to obtain a TAAL using anisotropic plies whose elastic properties  satisfy  conditions (\ref{eq:setAB}) or, alternatively, (\ref{eq:setABalphabeta}),  (\ref{eq:condsetAB}), (\ref{eq:condEiGij}), (\ref{eq:boundsgammaalpha}) and (\ref{boundsalphaGamma}).

\subsubsection{Existence of classical anisotropic plies for  fabricating TAALs}
\begin{figure}
\centering
\includegraphics[width=0.4\textwidth]{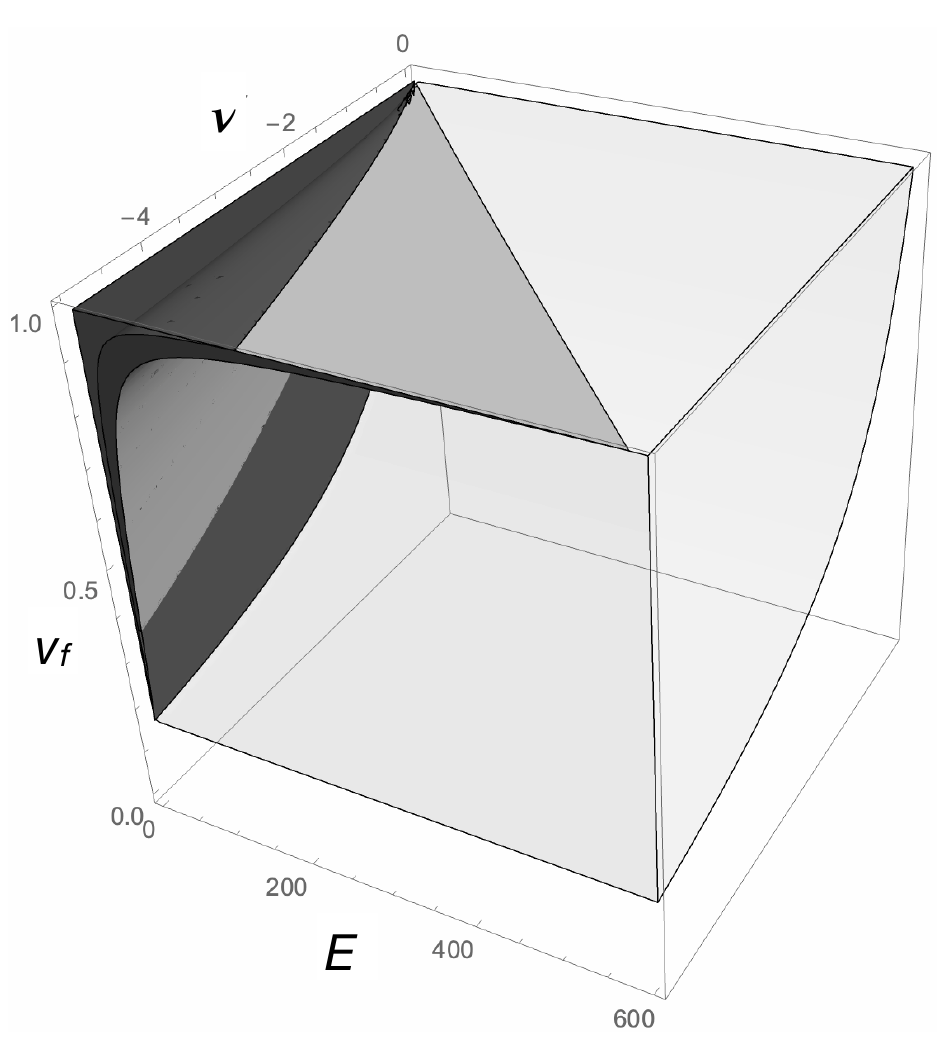}
\caption{Domain of existence of plies with $R_1=0$ suitable for the fabrication of TAALs or PAALs; the darker part corresponds to points of subset A, the gray one to points of subset B and the lighter gray to points of subset C.}
\label{fig:8}
\end{figure}

A final step is to evaluate whether or not it is possible to satisfy these conditions using anisotropic layers that do not have a microstructure giving rise to auxeticity. In other words, it is interesting to analyse whether or not it is possible to fabricate TAALs using anisotropic plies {\it à la Cauchy} reinforced by balanced fabrics. To this end, a homogenization law giving the elastic properties of a layer fabricated using a matrix and a balanced fabric must be used. Such a law depends upon the way the fabric is realized. Using the homogenization law given in \cite{gay14}, it is
\be
\label{eq:homogsq}
\begin{array}{l}
E_1=f_r\ E_\ell+(1-f_r)E_t,\medskip\\
E_2=(1-f_r)E_\ell+f_r\ E_t,\medskip\\
G_{12}=G_{\ell t},\medskip\\
\nud=\dfrac{\nu_{\ell t}E_t}{f_r\ E_t+(1-f_r)E_\ell},
\end{array}
\ee
where $f_r$ is the fibre's ratio defined by
\be
f_r:=\frac{f_1}{f_1+f_2},
\ee
with $f_1$ and $f_2$ respectively the number of fibres in the directions $x_1$ and $x_2$ of the layer.  $E_1,E_2,G_{12}$ and $\nud$ are the engineering constants of the layer, while $E_\ell, E_t,G_{\ell t}$ and $\nu_{\ell t}$ are those of a unidirectional layer, $\ell$ and $t$ stand for longitudinal and transverse, respectively. The layer is considered as composed of two sub-layers, one with $f_1$ fibres aligned in the direction $x_1$ and the other one with $f_2$ fibres along $x_2$. The constants $E_\ell, E_t,G_{\ell t}$ and $\nu_{\ell t}$ are determined using the rule of mixtures, \cite{jones, gay14}. In order to reduce the dimension of the problem, it is worth also in this case to use dimensionless quantities. To this end, the following parameters are defined:
\be
\label{eq:adimparameters}
\begin{array}{l}
E:=\dfrac{E_f}{E_m},\  \ \nu:=\dfrac{\nu_f}{\nu_m},\ \ G:=\dfrac{G_f}{G_m}=\dfrac{1+\nu_m}{1+\nu\ \nu_m}E,\medskip\\
\wE_\ell:=\dfrac{E_\ell}{E_m},\  \ \wE_t:=\dfrac{E_t}{E_m},\ \ \wnu_{\ell t}:=\dfrac{\nu_{\ell t}}{\nu_m},\ \ \wG_{\ell t}:=\dfrac{G_{\ell t}}{E_m},\medskip\\
\wE_1:=\dfrac{E_1}{E_m},\ \ \wE_2:=\dfrac{E_2}{E_m},\ \ \wnu_{12}:=\dfrac{\nud}{\nu_m},\ \ \wG_{12}:=\dfrac{G_{12}}{E_m},
\end{array}
\ee
with $E_f,E_m$ and $\nu_f,\nu_m$ the Young's moduli and the Poisson's ratios for the fibres and the matrix, respectively. Because fibres are intended to reinforce the matrix, i.e. $E_f>E_m$, and considering isotropic matrix and fibres, then
\be
E>1,\ \ -\frac{1}{\nu_m}<\nu<\frac{1}{2\nu_m}.
\ee
Then, recalling that 
\be
\frac{G_f}{E_m}=\frac{1}{E_m}\frac{E_f}{2(1+\nu_f)}=\frac{E}{2(1+\nu\ \nu_m)},
\ee
the rule of mixtures gives
\be
\label{eq:romadim}
\begin{array}{l}
\wE_\ell=1+\vf\ (E-1),\medskip\\
\wE_t=\dfrac{E}{\vf+(1-\vf)E},\medskip\\ 
\wnu_{\ell t}=1+\vf\ (\nu-1),\medskip\\
\wG_{\ell t}=\dfrac{E}{2[\vf\ (1+\nu\ \nu_m)+(1-\vf)(1+\nu_m)E]},
\end{array}
\ee
with $\vf$ the volume fraction of the fibres, $\vf\in[0,1]$. Because for a balanced fabric $f_r=1/2$, the dimensionless form of eq. (\ref{eq:homogsq}) is
\be
\label{eq:adimtech}
\begin{array}{l}
\wE_1=\wE_2=\dfrac{E+[1+\vf\ (E-1)][\vf+(1-\vf)E]}{2[\vf+(1-\vf)E]},\medskip\\
\wnu_{12}=\dfrac{2E[1+\vf\ (\nu-1)]}{E+[1+\vf\ (E-1)][\vf+(1-\vf)E]},\medskip\\
\wG_{12}=\dfrac{E}{2[\vf\ (1+\nu\ \nu_m)+(1-\vf)(1+\nu_m)E]}.
\end{array}
\ee
Also conditions (\ref{eq:condEiGij})  can be written in the form:
\be
\label{eq:adimboundsAB}
\begin{array}{l}
- \mathrm{for\ set\ A:}\ \ \ \wG_{12}>\dfrac{\wE_1}{2(1-\nu_m\wnu_{12})},\ \ \wnu_{12}<0,\medskip\\
- \mathrm{for\ set\ B:}\ \ \ \wG_{12}>\dfrac{1+3\nu_m\wnu_{12}}{2(1-\nu_m^2\wnu_{12}^2)}\wE_1\ \mathrm{and}\ \left\{\wG_{12}<\dfrac{\wE_1}{2(1-\nu_m\wnu_{12})}\ \mathrm{or}\ \wnu_{12}>0\right\}.
\end{array}
\ee
Injecting quantities in eq. (\ref{eq:adimtech})  into eq. (\ref{eq:adimboundsAB}) gives the conditions for the existence of TAALs. These conditions are not written here because too much long; their analytical solution is not possible, but the graphical representation of the existence domain is given in Fig. \ref{fig:8} for the case of a matrix with $\nu_m=0.25$ (the influence of this parameter is very low, the domain does not change substantially for other values of $\nu_m$).
From Fig. \ref{fig:8} it can be seen that a solution can exist only for $\nu<0$, i.e. for layers composed of two phases whose at least one is by itself auxetic. In other words, it is impossible to realize TAALs using non-auxetic materials: total auxeticity cannot be get as an effect  of anisotropy uniquely. This result conforms what already found for TAALs composed of unidirectional layers, cf. \cite{vannucci24b}. Also, it can be seen that actually the conditions on $\wnu_{12}$ in eq. (\ref{eq:adimboundsAB}) are inessential: the domain of existence of TAALs is hence simply defined by the bounds
\be
\begin{array}{l}
- \textrm{for set A :}\ \ \ \wG_{12}>\dfrac{\wE_1}{2(1-\nu_m\wnu_{12})},\medskip\\
- \textrm{for set B :}\ \ \ \dfrac{1+3\nu_m\wnu_{12}}{2(1-\nu_m^2\wnu_{12}^2)}\wE_1<\wG_{12}<\dfrac{\wE_1}{2(1-\nu_m\wnu_{12})}.
\end{array}
\ee

\subsection{Partially Auxetic Anisotropic Laminates (PAALs)}
We ponder now on the existence of Partially  Auxetic Anisotropic Laminates. The procedure is quite similar to the previous one used to study TAALs, the same steps are followed.
\subsubsection{Polar conditions}
To obtain a PAAL, eq. (\ref{eq:aux3}) must be satisfied for at least one $\theta$. Because $\mu(\xi_1,\theta)$ gets its maximum  for $\theta=0$, the condition for having a TAAL is
\be
\tilde{\mu}(\xi_1)=2\tau-1+\rho^2\xi_1^2-2\tau \rho\ \xi_1<0,\ \ \ \xi_1\in[0,1].
\ee
Proceeding like in the case of TAALs,
the bounds of the admissible set given by the above inequality are
\be
(\xi_1)_{1,2}=\frac{\tau\pm|\tau-1|}{\rho}. 
\ee
Once more, remembering  that 
$\xi_1\in[0,1]$ and eq. (\ref{eq:elboundpoladim}), by some standard steps it can be checked that PAALs are possible only for 
\be
\xi_1\in\left[\dfrac{2\tau-1}{\rho},1\right]
\ee
when
\be
\label{eq:setPAAL}
\dfrac{1}{2}<\tau<\dfrac{1+\rho}{2}.
\ee
This subset of admissible points $(\tau,\rho)$ for PAALs will be  denoted by C, see Fig. \ref{fig:1}. It is evident by itself that it is meaningless to speak of PAALs $\forall\xi_1$. 

The above conditions can be given in a dimensional form coming back to polar parameters $T_0,T_1,R_0$: 
\be
\label{eq:PAAL1}
\frac{T_0}{2}<T_1<\frac{T_0+R_0}{2};
\ee
to remember the general elastic bounds (\ref{eq:elboundpol}), in particular $0<R_0<T_0$. 
It is to be noticed that also points of subset B can give rise to PAALs, like the example in Fig. \ref{fig:6} shows in the particular case of just one layer. However, such points can give rise also to TAALs, while those of subset C only to PAALs. That is why in the following we will continue to analyze points of subset C.

\subsubsection{Cartesian conditions}
Using again eq. (\ref{eq:polcart}),  conditions (\ref{eq:PAAL1}) can be given as functions of the Cartesian components $\Q_{ij}$. If once more $\Q_{11}=\Q_{22}$ is taken into account, then bounds (\ref{eq:PAAL1})  are equivalent to
\be
\label{eq:PAAL3}
\Q_{66}<\Q_{11}+3\Q_{12}<\Q_{66}+ |\Q_{11}-\Q_{12}-\Q_{66}|.
\ee
Passing again to parameters $\alpha$ and $\beta$, the above relations become the bounds
\be
\label{eq:PAAL2}
\beta<1+3\alpha<\beta+|1-\alpha-\beta|,
\ee
that define the domain C shown in Fig. \ref{fig:2}. 

\subsubsection{Technical moduli conditions}
If eq. (\ref{eq:Qijs}) is injected into eq. (\ref{eq:PAAL3}) we get
\be
\label{eq:paalboundmoduli}
G_{12}<\frac{1+3\nud}{2(1-\nud^2)}E_1<G_{12}+\left|\frac{E_1}{2(1+\nud)}-G_{12}\right|.
\ee
Passing again to parameters $\alpha$ and $\gamma$, the above bounds become
\be
\gamma<\frac{1+3\alpha}{2(1-\alpha^2)}<\gamma+\left|\frac{1}{2(1+\alpha)}-\gamma\right|,
\ee
that define the subset C in Fig. \ref{fig:3}. If instead parameter $\Gamma$ is used in place of $\gamma$, then it is  easily get  
\be
\Gamma<\frac{1+3\alpha}{1-\alpha}<\Gamma+|1-\Gamma|,
\ee
which gives the domain C in Fig. \ref{fig:4}.

\subsubsection{Further considerations and numerical examples}
\begin{figure}[h]
\begin{center}
\begin{tikzpicture}
\begin{polaraxis} [width=0.5\textwidth]
 \addplot[ultra thick,domain=0:360,samples=360]
{0.001/(0.0000483059+0.00000824177*cos(4*x))])};
\addplot[thick,domain=0:360,samples=360]
{0.001/(4*(0.0000274725-0.00000824177*cos(4*x)))};
\end{polaraxis}
\end{tikzpicture}
\begin{tikzpicture}
\begin{polaraxis} [width=0.5\textwidth,ytick={-0.5}]
 \addplot[ultra thick,domain=0:360,samples=360]
{0.3+(0.0000066391+0.00000824177*cos(4*x))/(0.0000483059 +0.00000824177*cos(4*x))};
\addplot[ domain=0:360,samples=360]
{0.3};
\end{polaraxis}
\end{tikzpicture}
\caption{Polar diagrams for the layer of Example 3; left: $E_1(\theta),G_{12}(\theta)$, respectively thick and thin curves; right:  $\nu_{12}(\theta)$ (the thin circular line corresponds to zero: inside it, $\nud<0$).}
\label{fig:9}
\end{center}
\end{figure}
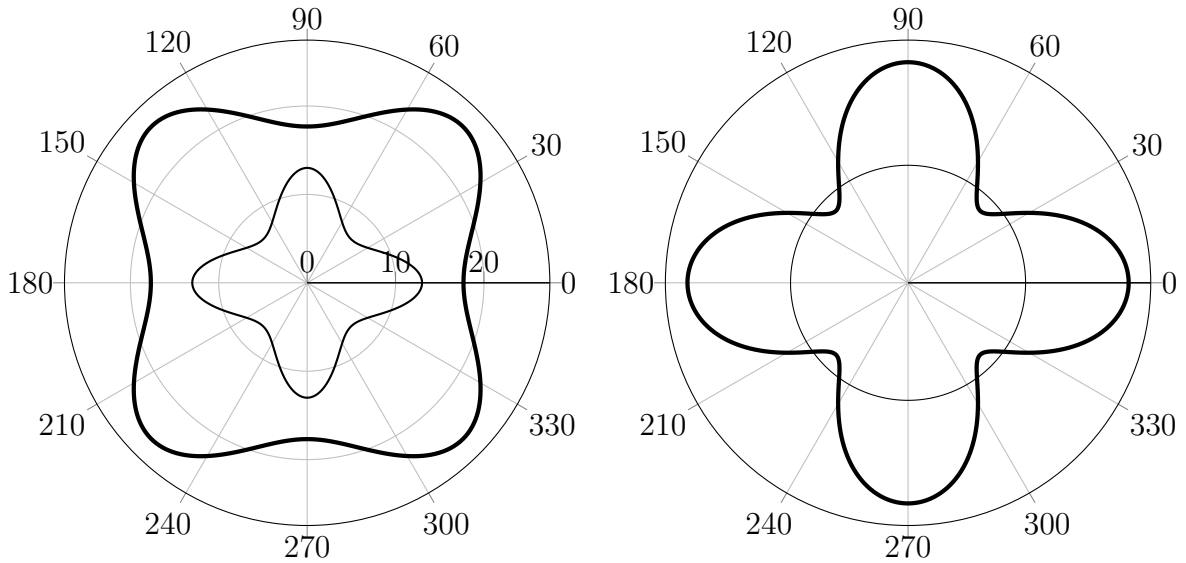

{\bf Example 3}: as first example of PAALs, we consider a material with $E_1=E_2=17.68, G_{12}=13,\nu_{12}=0.263\Rightarrow\Q_{11}=\Q_{22}=19,\Q_{12}=5,\Q_{66}=26$, which gives $T_0=10,T_1=6,R_0=3,\Phi_0=45^\circ\Rightarrow\tau=0.6,\rho=0.3$, so a point of set C, indicated by label "3" in Fig. \ref{fig:1}. The polar diagrams of $E_1(\theta),G_{12}(\theta)$ and $\nu(\theta)$ are shown in Fig. \ref{fig:9}. 

As apparent from the diagram of $\nud(\theta)$, the same layer is partially auxetic. The question is hence: is it possible to obtain a PAAL using a layer that is not at all auxetic, i.e. having $\nud(\theta)>0\ \forall\theta$? For such a situation, it should be 
\be
\min_\theta\mu(\xi_1,\theta)>0.
\ee
 If a laminate is composed by just one layer, orienting it at $0^\circ$ then $\xi_1=1$, cf. eq. (\ref{eq:lampar}). So the condition for having $\nud(\theta)>0\ \forall\theta$ is
 \be
 \min_\theta\mu(1,\theta)=\mu(1,0)=\rho^2-2\tau\rho+2\tau-1>0,
 \ee
condition which is satisfied for
\be
\frac{1+\rho}{2}<\tau<1,
\ee
i.e. by points that are out of the subset C. This implies that it is impossible to realize PAALs using square symmetric  layers completely non auxetic or, in other words, that only partially auxetic square symmetric layers can produce PAALs. This is a substantial difference with respect to unidirectional layers, that can produce PAALs also when they are completely non auxetic, cf. \cite{vannucci24b}.

Another difference with respect to unidirectional plies, is that now it is possible to determine analytically the lowest possible value of $\nud^\A(\theta)$, which can be done only numerically for unidirectional layers, cf. \cite{vannucci24d}. Putting equal to zero the gradient of $\nud^\A(\xi_1,\theta)$ in the set $\xi_1\in[0,1],\theta\in[0,\pi/4]$ gives that the stationary point is
\be
({\nud^\A})_1=\frac{2\tau-1}{2\tau+1}\ \ \ \mathrm{for}\ \ \xi_1=0,\ \theta=\frac{\pi}{8}.
\ee
Actually, $\xi_1=0$ corresponds to an isotropic laminate, cf. \cite{CompStruct02}. 
However, the minimum can be get also on the boundary, i.e. for $\xi_1=1$, where the stationary points  of $\nud^\A$ are: 
\be
({\nud^\A})_2=\frac{2\tau-1+\rho^2-2\tau\rho}{2\tau+1-\rho^2-2\tau\rho} \ \ \ \mathrm{for}\ \ \theta=0;
\ee
\be
({\nud^\A})_{3}=\frac{2\tau-1+\rho^2+2\tau\rho}{2\tau+1-\rho^2+2\tau\rho}\ \ \ \mathrm{for}\ \ \theta=\frac{\pi}{4}.
\ee
It can be easily checked that in the set (\ref{eq:elboundpoladim}) the minimum is $({\nud^\A})_2$, which is get for $\xi_1=1,\theta=0$, i.e. for the same conditions giving the minimum of $\nud(\theta)$ of a single ply. 
This means that superposing partially auxetic layers will give rise to a PAAL whose minimum $\nud^\A(\theta)$ will never be lower than the minimum of $\nud(\theta)$ of the single layer. 

The only possible design problem concerning PAALs is hence the maximization of the auxetic zone. To this end, cf. \cite{vannucci24d}, the zone where $\nud^\A(\theta)<0$ is determined by the condition
\be
\cos4\theta>\lambda(\xi_1):=\frac{2\tau-1+\rho^2\xi_1^2}{2\tau\rho\xi_1},
\ee
so the largest auxetic zone will be obtained minimizing function $\lambda(\xi_1)$, with the obvious condition $|\lambda|\leq1$. 
Studying the variation of $\lambda(\xi_1)$ by elementary methods shows that:
\begin{enumerate}[i.]
\item for $\dfrac{1-\rho}{2}<\tau<\dfrac{1}{2}$, i.e. for points of subset B, 
\be
\lambda_{\min}=-1 \ \ \ \mathrm{for}\ \ \ \xi_1=\frac{2\tau-1}{\rho};
\ee
because $\cos4\theta\geq-1$, it is obvious that in this case $\nud^\A(\theta)<0\ \forall\theta$, and in fact for points of subset B,  TAALs are possible;
\item for $\dfrac{1}{2}<\tau<\dfrac{1+\rho}{2}$, i.e. for points of subset C,
\be
\begin{array}{l}
\lambda_{min}=\dfrac{\sqrt{2\tau-1}}{\tau}\ \ \ \mathrm{for}\ \ \ \xi_1=\dfrac{\sqrt{2\tau-1}}{\rho}\ \ \ \mathrm{if}\ \ \ \rho>\sqrt{2\tau-1};\medskip\\
\lambda_{min}=1\ \ \ \mathrm{for}\ \ \ \xi_1=1\ \ \ \mathrm{if}\ \ \ \rho<\sqrt{2\tau-1}.
\end{array}
\ee
\end{enumerate}
The curve $\rho=\sqrt{2\tau-1}$ is shown in Fig. \ref{fig:1}. The above result indicates that only for points of set C above this curve it is possible to optimize the auxetic zone, while for those beneath the curve, the minimum is get for $\xi_1=1$, i.e. for the single layer: no improvement of the auxetic zone can be obtained by a laminate realized with such a layer, which is also the case of material of Example 3.

\begin{figure}[h]
\begin{center}
\begin{tikzpicture}
\begin{polaraxis} [width=0.5\textwidth]
 \addplot[ultra thick,domain=0:360,samples=360]
{0.001/(0.0000560605+0.000016667*cos(4*x))])};
\addplot[thick,domain=0:360,samples=360]
{0.001/(4*(0.000033333-0.0000166666*cos(4*x)))};
\end{polaraxis}
\end{tikzpicture}
\begin{tikzpicture}
\begin{polaraxis} [width=0.5\textwidth,ytick={-0.5}]
 \addplot[ultra thick,domain=0:360,samples=360]
{0.3+(0.0000106061+0.0000166667*cos(4*x))/(0.0000560605 +0.0000166667*cos(4*x))};
\addplot[ domain=0:360,samples=360]{0.3};
\end{polaraxis}
\end{tikzpicture}
\caption{Polar diagrams for the layer of Example 4; left: $E_1(\theta),G_{12}(\theta)$, respectively thick and thin curves; right:  $\nu_{12}(\theta)$ (the thin circular line corresponds to zero: inside it, $\nud<0$).}
\label{fig:10}
\end{center}
\end{figure}
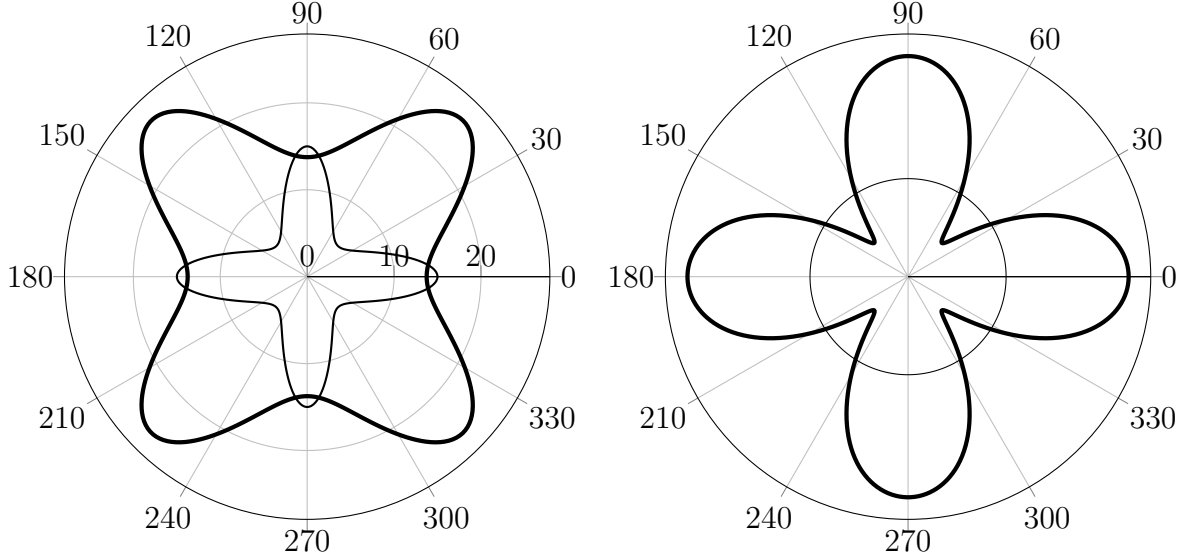

\begin{figure}[h]
\begin{center}
\begin{tikzpicture}
\begin{polaraxis} [width=0.48\textwidth,ytick={-0.5}]
 \addplot[ thick,domain=0:360,samples=360]
{0.3+(0.0000050464+0.00000877687*cos(4*x))/(0.0000505008+0.00000877687*cos(4*x))};
\addplot[ thick,densely dashed,domain=0:360,samples=360]
{0.3+(0.0000106061+0.0000166667*cos(4*x))/(0.0000560605 +0.0000166667*cos(4*x))};
\addplot[ domain=0:360,samples=360]{0.3};
\end{polaraxis}
\end{tikzpicture}
\begin{tikzpicture}
\begin{axis} [width=0.48\textwidth,height=0.5\textwidth,xtick={0,15,30,45,60,75,90},xmin=0,xmax=90,ylabel=${\nud,\nud^\A}$]
 \addplot[ thick,domain=0:90,samples=90]
{(0.0000050464+0.00000877687*cos(4*x))/(0.0000505008+0.00000877687*cos(4*x))};
\addplot[ thick,densely dashed,domain=0:90,samples=90]
{(0.0000106061+0.0000166667*cos(4*x))/(0.0000560605 +0.0000166667*cos(4*x))};
\addplot[domain=0:90]{0};
\end{axis}
\end{tikzpicture}
\caption{Polar and Cartesian diagrams of $\nud(\theta)$, dashed line, and of $\nud^\A(\theta)$, solid line, for the  material of Example 4 and the laminate  with $\xi_1=0.632$ made with this material (the thin circular line corresponds to zero: inside it, $\nud<0$). }
\label{fig:12}
\end{center}
\end{figure}
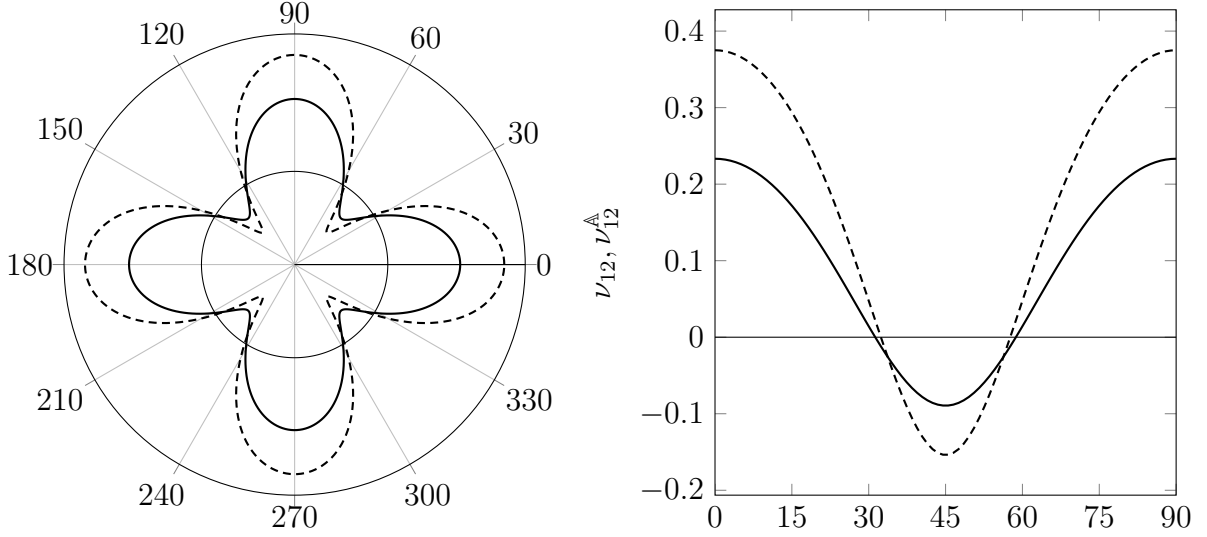

{\bf Example 4}: as second example of PAALs, we consider a material with $E_1=E_2=13.75, G_{12}=15,\nu_{12}=0.375\Rightarrow\Q_{11}=\Q_{22}=16,\Q_{12}=6,\Q_{66}=30$, which gives $T_0=10,T_1=5.5,R_0=5,\Phi_0=45^\circ\Rightarrow\tau=0.55,\rho=0.5$, so a point of set C, indicated by label "4" in Fig. \ref{fig:1}. The polar diagrams of $E_1(\theta),G_{12}(\theta)$ and $\nu(\theta)$ are shown in Fig. \ref{fig:10}. The amplitude $\Delta\alpha$ of the auxetic zone is
\be
\Delta\alpha=\frac{1}{2}\arccos(\lambda(\xi_1=1))=\frac{1}{2}\arccos\left(\frac{2\tau-1+\rho^2}{2\tau\rho}\right)=25.2^\circ.
\ee
It is possible for such a material to maximize the auxetic zone; in particular, this happens for $\xi_1=0.632$, e.g. for any  angle-ply laminate with orientations
\be
\delta=\pm\frac{1}{4}\arccos(\xi_1)=\pm12.7^\circ.
\ee
For such a case the amplitude $\Delta\alpha$ of the auxetic zone is
\be
\Delta\alpha=\frac{1}{2}\arccos(\lambda_{min})=\frac{1}{2}\arccos\left(\dfrac{\sqrt{2\tau-1}}{\tau}\right)=27.5^\circ,
\ee
with an increment of $2.3^\circ$, i.e. of about 9\% with respect to single layer. 
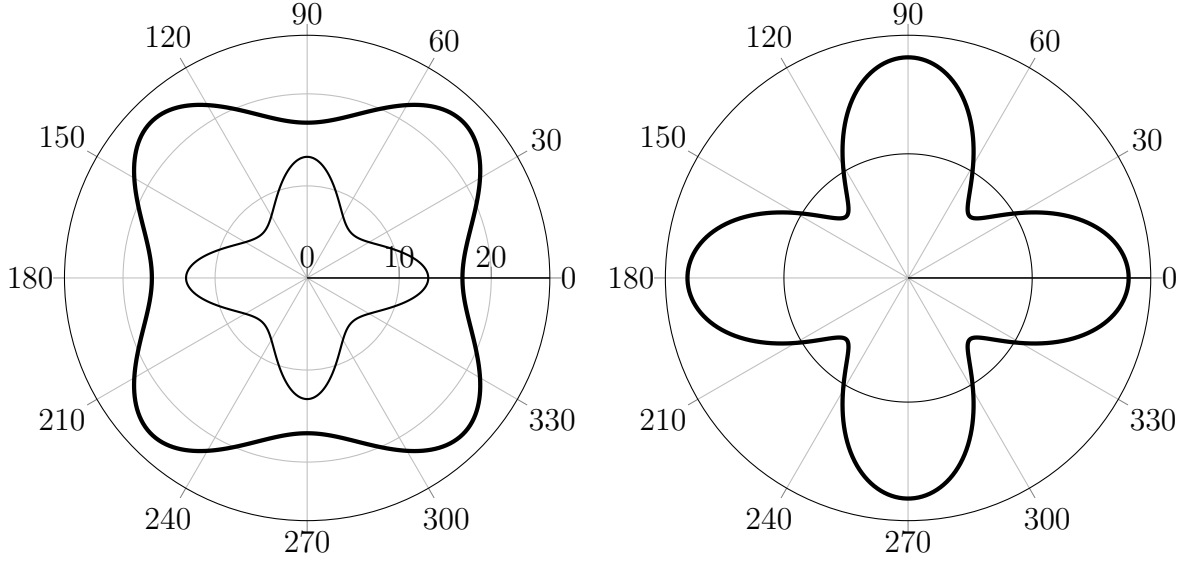
\begin{figure}[h]
\begin{center}
\begin{tikzpicture}
\begin{polaraxis} [width=0.5\textwidth]
 \addplot[ultra thick,domain=0:360,samples=360]
{0.001/(0.0000505008+0.00000877687*cos(4*x))])};
\addplot[thick,domain=0:360,samples=360]
{0.001/(4*(0.0000277736 -0.00000877687*cos(4*x)))};
\end{polaraxis}
\end{tikzpicture}
\begin{tikzpicture}
\begin{polaraxis} [width=0.5\textwidth,ytick={-0.5}]
 \addplot[ultra thick,domain=0:360,samples=360]
{0.3+(0.0000050464+0.00000877687*cos(4*x))/(0.0000505008+0.00000877687*cos(4*x))};
\addplot[ domain=0:360,samples=360]{0.3};
\end{polaraxis}
\end{tikzpicture}
\caption{Polar diagrams for the laminate with $\xi_1=0.632$ made with the material of Example 4; left: $E_1^\A(\theta),G_{12}^\A(\theta)$, respectively thick and thin curves; right:  $\nu_{12}^\A(\theta)$ (the thin circular line corresponds to zero: inside it, $\nud^\A<0$).}
\label{fig:11}
\end{center}
\end{figure}
The polar diagrams of $E_1^\A(\theta),G_{12}^\A(\theta)$ and $\nud^\A(\theta)$ for such a laminate are shown in Fig. \ref{fig:11}, while the diagrams of $\nud(\theta)$ and of $\nud^\A(\theta)$ are compared in Fig. \ref{fig:12}.
It is apparent  the increase of the auxetic zone, while, as shown above, the minimum negative value is higher for the laminate :
\be
{\nud}_{min}=-0.15 <{\nud^\A}_{min}=-0.09.
\ee

\subsubsection{Existence of classical anisotropic plies for fabricating PAALs}
The last question concerns once more the possible existence of plies composed by non auxetic materials able to realize PAALs. The procedure is quite the same already used for TAALs, with now the dimensionless corresponding  of bounds (\ref{eq:paalboundmoduli}) that are
\be
\wG_{12}<\frac{1+3\nu_m\wnu_{12}}{2(1-\nu_m^2\wnu_{12}^2)}E\ \ \mathrm{and}\ \ \wnu_{12}<0 \ \ \mathrm{or}\ \ \frac{E}{2(1-\nu_m\wnu_{12})}<\wG_{12}<\frac{1+3\nu_m\wnu_{12}}{2(1-\nu_m^2\wnu_{12}^2)}E.
\ee
Once more, if now relations in eq. (\ref{eq:adimtech}) are used, the domain of parameters $E,\nu, \vf$ to fabricate plies able to realize PAALs is found. It is represented in Fig. \ref{fig:8} and also in this case it can be observed that materials with a negative Poisson's ratio must be used ($\nu<0$ everywhere).

This result is another difference with respect to unidirectional plies: it is impossible to obtain auxeticity as a simple result of anisotropy when plies reinforced by balanced fabrics are used. It is necessary to use either a matrix or fibres that are auxetic, which can be obtained only by some kind of microstructure, i.e. not using classical {\it à la Cauchy} non auxetic materials. The high degree of anisotropy given to a ply by a unidirectional set of fibres seems hence to be a necessary, though not sufficient, condition to obtain auxeticity in composite materials as a simple fact of anisotropy, see discussion in  \cite{vannucci24b}.

Of course, the above results, just like those found for TAALs, are coherent with the homogenization (\ref{eq:homogsq}) and other homogenization criteria can lead to different results. For instance, if a homogenization based upon the classical laminated plates theory model is used, where the warp and weft fibres are consider to belong to two identical layers superposed and rotated of $90^\circ$, then the relations (\ref{eq:homogsq}) will be different. However a test made with such a theory shows that the results are substantially the same, i.e., in particular, that  solutions are possible only for layers with $\nu<0$.

\section{Laminates composed of $R_0-$orthotropic layers}

\subsection{Dimensionless auxeticity condition}
Let us now turn the attention to the case of laminates composed of $R_0-$orthotropic identical layers. We follow the same steps of the previous case. First of all, $R_0=0\Rightarrow R_0^\A=0$; then, multiplying eq. (\ref{eq:homoga})$_4$ by $\mathrm{e}^{-4\mathrm{i}\Phi_0}$ we get
\be
R_1^\A\mathrm{e}^{4\mathrm{i}(\Phi_1^\A-\Phi_1)}=R_1(\xi_3+\mathrm{i}\xi_4),
\ee
so choosing
\be
\Phi_1^\A-\Phi_1=c_1\pi,\ \ \ c_1\in\mathbb{N},
\ee
i.e., by a proper choice of the reference frames of the layer and of the laminate, it is
\be
\label{eq:R1laminato}
\xi_4=0\ \Rightarrow\ R_1^\A=R_1\xi_3,
\ee
with 
\be
\xi_3=\frac{1}{n}\sum_{j=1}^n\cos2\delta_j,\ \ \ \xi_3\in[-1,1].
\ee
Fixing, as normally done, $\Phi_1^\A=0$, the auxeticity condition for the laminate, eq. (\ref{eq:aux1}), becomes now
\be
2(T_0T_1-R_1^2\xi_3^2)-T_0^2+2R_1^2\xi_3^2\cos4\theta<0.
\ee
Once more, for the sake of simplicity, we observe that it is sufficient to take into consideration just the laminates with $\xi_3\in[0,1]$, because those with $\xi_3\in[-1,0]$ can be obtained by a simple rotation of $\pi/2$; moreover, we introduce again the dimensionless parameter $\tau$ while $\rho$ is now replaced by
\be
\sigma:=\frac{R_1}{T_0}.
\ee
The elastic bounds for a $R_0-$orthotropic ply are, see \cite{vannucci_libro},
\be
T_0T_1>2R_1^2,\ \ \ R_1>0,
\ee
i.e., in dimensionless form,
\be
\label{eq:boundsR0adim}
\tau>2\sigma^2,\ \ \ \sigma>0.
\ee
Finally, the auxeticity dimensionless condition is 
\be
\label{eq:auxeticR0cond}
\eta(\xi_3,\theta):=2\tau-2\sigma^2\xi_3^2(1-\cos4\theta)-1<0.
\ee

\subsection{Totally Auxetic Anisotropic Laminates (TAALs)}
\subsubsection{Polar conditions}
\begin{figure}
\begin{center}
\begin{tikzpicture}
\begin{axis} [xmin=0,xmax=1.,xlabel=$\sigma$,
 ymin=0,ymax=1.5,ylabel=$\tau$, width=5cm,height=7.5cm,scale only axis,samples=100]
\addplot[ name path=A,domain=0:2]{0};
\addplot[ thick,name path=B,domain=0:0.5]{1/2};
\addplot[ thick,name path=C,domain=0:2]{1/2+2*x^2};
\addplot[ thick,name path=D,domain=0:0.5]{2*x^2};
\addplot[ thick,name path=E,domain=0:2]{2*x^2};
\addplot[gray,fill opacity=0.5,domain=0:0.5] fill between[of=D  and B] ;
\addplot[gray,fill opacity=0.1] fill between[of=E  and C] ;
\addplot [only marks,domain=0:1] coordinates{(0.2,0.2)(0.4,0.6)(0.648,0.7)};
\pgfplotsset{every axis/.append style={extra description/.code={\node at (0.2,0.25) {\Large{T}};\node at (0.55,0.6){\Large{P}};\node at (0.2,0.17){$5$};\node at (0.4,0.435){$6$};\node at (0.648,0.50){$7$};}}}
\end{axis}
\end{tikzpicture}
\caption{Domains of solutions for TAALs, subset T, and PAALs, subset P, composed of $R_0=0$ layers, in the plane $(\sigma,\tau)$.}
\label{fig:13}
\end{center}
\end{figure}
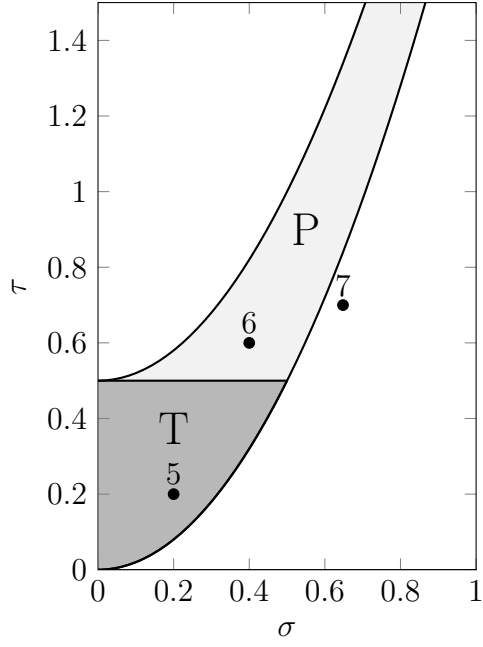

The last equation must be satisfied $\forall\theta$ to fabricate a TAAL. This happens $\iff\theta=k\pi/2$, $ k\in\mathbb{N}$, which finally gives the condition
\be
\hat{\eta}(\xi_3)=2\tau-1<0.
\ee
This inequality does not depend upon $\xi_3$, i.e. on the stacking sequence. As a consequence, and rather surprisingly unlike unidirectional or square-symmetric plies, the possibility of fabricating a TAAL depends uniquely on the properties of the ply, which  must be itself a TAAL, and not also on the stacking sequence, i.e. on $\xi_3$. Finally, taking into account the elastic bounds 
(\ref{eq:boundsR0adim}), the subset in the space $(\sigma,\tau)$ for the existence of TAALS composed of $R_0-$orthotropic plies is defined by the bounds
\be
\label{eq:existtaalR0}
2\sigma^2<\tau<\frac{1}{2}
\ee
and indicated in Fig. \ref{fig:13} by the label T.
Coming back to polar parameters, the above bounds read like
\be
2\frac{R_1^2}{T_0}<T_1<\frac{T_0}{2}.
\ee
\subsubsection{Cartesian conditions}
Using eq. (\ref{eq:polcart}), the above bounds become
\be
\label{eq:bornesR0cart}
\begin{array}{l}
 \Q_{11}^2-6\Q_{11}\Q_{22}+\Q_{22}^2+4\Q_{12}^2-2\Q_{66}(\Q_{11}+2\Q_{12}+\Q_{22})<0,\medskip\\
 \Q_{11}+6\Q_{12}+\Q_{22}-2\Q_{66}<0.
 \end{array}
\ee
The comparison of these bounds with the previous ones shows, once more, how much  the polar formalism is more effective than the Cartesian one in defining the anisotropy conditions in plane elasticity.

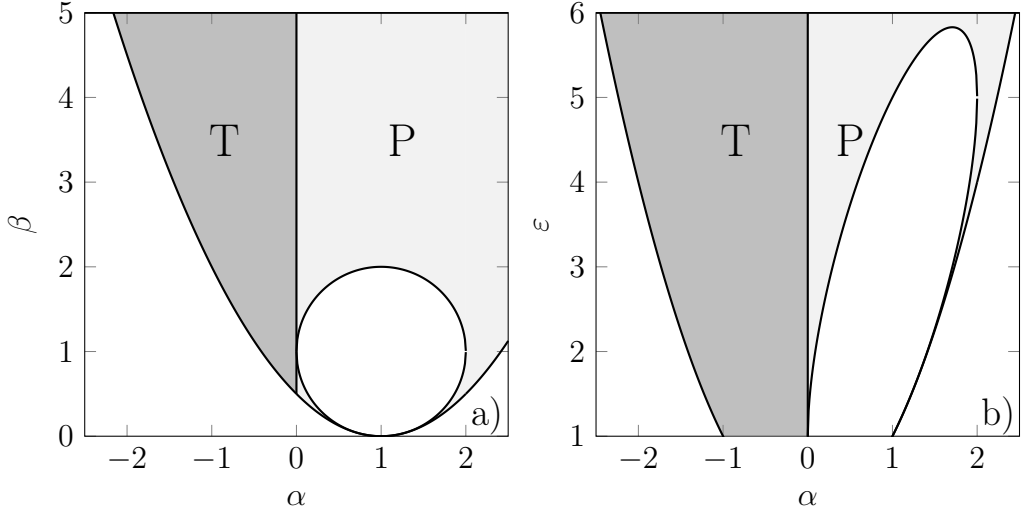
\begin{figure}
\begin{center}
\begin{tikzpicture}
\begin{axis} [xmin=-2.5,xmax=2.5,xlabel=$\alpha$,
 ymin=0,ymax=5,ylabel=$\beta$, width=.35\textwidth,height=.35\textwidth,samples=200,scale only axis]
\addplot[ thick,name path=A,domain=-2.5:0]{((x-1)^2)/2};
\addplot[ thick,name path=B,domain=-2.5:0]{5};
\addplot[thick]coordinates{(0,0.5)(0,5)};
\addplot[ thick,name path=C,domain=0:2]{((x-1)^2)/2};
\addplot[ thick,name path=D,domain=2:2.5]{((x-1)^2)/2};
\addplot[thick,name path=E,domain=0:2]{1-(2*x-x^2)^0.5};
\addplot[thick,name path=F,domain=0:2]{1+(2*x-x^2)^0.5};
\addplot[ thick,name path=G,domain=0:2]{5};
\addplot[ thick,name path=H,domain=2:2.5]{5};
\addplot[gray,fill opacity=0.5] fill between[of=A  and B] ;
\addplot[gray,fill opacity=0.1] fill between[of=C  and E] ;
\addplot[gray,fill opacity=0.1] fill between[of=F  and G] ;
\addplot[gray,fill opacity=0.1] fill between[of=D  and H] ;
\pgfplotsset{every axis/.append style={extra description/.code={\node at (0.33,0.7) {\Large{T}};\node at (0.75,0.7) {\Large{P}};\node at (0.95,0.05){\large{a)}};}}}
\end{axis}
\end{tikzpicture}
\begin{tikzpicture}
\begin{axis} [xmin=-2.5,xmax=2.5,xlabel=$\alpha$,
 ymin=1,ymax=6,ylabel=$\eps$, width=.35\textwidth,height=.35\textwidth,samples=200,scale only axis]
\addplot[ thick,name path=A,domain=-2.5:0]{x^2};
\addplot[ thick,name path=B,domain=-2.5:0]{6};
\addplot[ thick,name path=C,domain=0:2]{x^2};
\addplot[thick,name path=D,domain=2:2.5]{x^2};
\addplot[ thick,name path=E,domain=0:2]{1+2*x-2*(2*x-x^2)^0.5};
\addplot[ thick,name path=F,domain=0:2]{1+2*x+2*(2*x-x^2)^0.5};
\addplot[ thick,name path=G,domain=0:2]{6};
\addplot[ thick,name path=H,domain=2:2.5]{6};
\addplot[thick]coordinates{(0,1)(0,6)};
\addplot[gray,fill opacity=0.5] fill between[of=A  and B] ;
\addplot[gray,fill opacity=0.1] fill between[of=C  and E] ;
\addplot[gray,fill opacity=0.1] fill between[of=F  and G] ;
\addplot[gray,fill opacity=0.1] fill between[of=D  and H] ;
\pgfplotsset{every axis/.append style={extra description/.code={\node at (0.33,0.7) {\Large{T}};\node at (0.6,0.7) {\Large{P}};\node at (0.95,0.05){\large{b)}};}}}
\end{axis}
\end{tikzpicture}
\caption{Domains of solutions for TAALs, subset T, and PAALs, subset P, composed of $R_0=0$ layers: a)  in the plane $(\alpha,\beta)$, b) in the plane $(\alpha,\eps)$.}
\label{fig:14}
\end{center}
\end{figure}

The last inequalities can be given in dimensionless form, using again parameters $\alpha$ and $\beta$. However now, because $\Q_{22}\neq\Q_{11}$, another dimensionless parameter is needed:
\be
\label{eq:epsdef}
\eps:=\frac{E_1}{E_2}=\frac{\Q_{11}}{\Q_{22}},\ \ \ \eps>1.
\ee
Actually, the special orthotropy $R_0$, introduces a relation among the elastic moduli (just like square symmetry, which in fact gives $\Q_{11}=\Q_{22}, E_1=E_2$). Such relation is particularly simple to be written with the dimensionless parameters:
\be
\label{eq:epsalfabeta}
\eps=2\alpha+2\beta-1.
\ee
This condition can be obtained either  using the definitions of $\alpha,\beta,\eps$ and expressing the $\Q_{ij}$s by the polar parameters $T_0,T_1,R_1$, then passing to $\tau$ and $\sigma$ and solving for $\eps$, or directly considering the Cartesian equivalent of the condition $R_0=0$, \cite{vannucci_libro}:
\be
\Q_{11}-2\Q_{12}-2\Q_{66}+\Q_{22}=0.
\ee
So, after some easy calculations, it is found that the dimensionless form of eq. (\ref{eq:bornesR0cart}) in the $(\alpha,\beta)$ space is
\be
\alpha<0,\ \ \ \beta>\frac{1}{2}(\alpha-1)^2,
\ee
subset which is labelled by T in Fig. \ref{fig:14} a). If now $\alpha$ and $\beta$ are replaced by their expressions in terms of the $\Q_{ij}$s, eq. (\ref{eq:alphabeta}), the last conditions become
\be
\Q_{12}<0,\ \ \ 2\Q_{22}\Q_{66}>(\Q_{12}-\Q_{22})^2,
\ee
totally equivalent to bounds in eq. (\ref{eq:bornesR0cart}) but much simpler.
\subsubsection{Technical moduli conditions}
Replacing eq. (\ref{eq:Qorthply}) into eq. (\ref{eq:bornesR0cart}) and remembering the reciprocity condition (\ref{eq:reciprocity}),  the conditions for obtaining TAALs in terms of the technical moduli of a $R_0$-orthotropic ply can be obtained:
\be
\label{eq:bornesTAALsR0consting}
\begin{array}{l}
E_1^3-2E_1^2(3E_2+2G_{12})+4E_2^2G_{12}\nud^2(1+2\nud)+\smallskip\\
+E_1E_2[E_2-4G_{12}(1+2\nud)+4(E_2+G_{12})\nud^2]<0,\bigskip\\
\dfrac{E_1(E_1+E_2-4G_{12})+6E_1E_2\nud+4E_2G_{12}\nud^2}{E_1-E_2\nud^2}<0.
\end{array}
\ee
 The above bounds can be given in dimensionless form. To this purpose,  the definitions of the dimensionless parameters $\gamma,\eps$, eqs. (\ref{eq:gammadef}), (\ref{eq:epsdef}) are used, along with the fact  that $\alpha=\nud$, to then observe that, through  eq. (\ref{eq:epsalfabeta}),
 \be
\beta=\frac{\Q_{66}}{\Q_{22}}=\frac{2G_{12}}{E_2}(1-\nud \nu_{12})=2\gamma\left(1-\frac{\alpha^2}{\eps}\right)\Rightarrow\gamma=\frac{\eps(\eps-2\alpha+1)}{4(\eps-\alpha^2)}.
\ee
Finally, some long but standard algebraic manipulations  give the dimensionless bounds  with only two variables, $\alpha$ and $\eps$, that are surprisingly simple :
\be
\label{eq:boundsmoduliadimR0ply}
\alpha<0,\ \ \ \eps>\alpha^2.
\ee
The corresponding subset of the space $(\alpha,\eps)$ is indicated by label T in Fig. \ref{fig:14} b). If one considers that $\alpha=\nud$ and $\eps=E_1/E_2$, the above bounds can be immediately rewritten  using the technical constants of the layer:
\be
\label{eq:boundsmoduliR0ply}
\nud<0,\ \ \ E_1>\nud^2E_2,
\ee
indeed very simple conditions completely equivalent to the bounds in eq. (\ref{eq:bornesTAALsR0consting}).

\subsubsection{Further considerations and  examples}
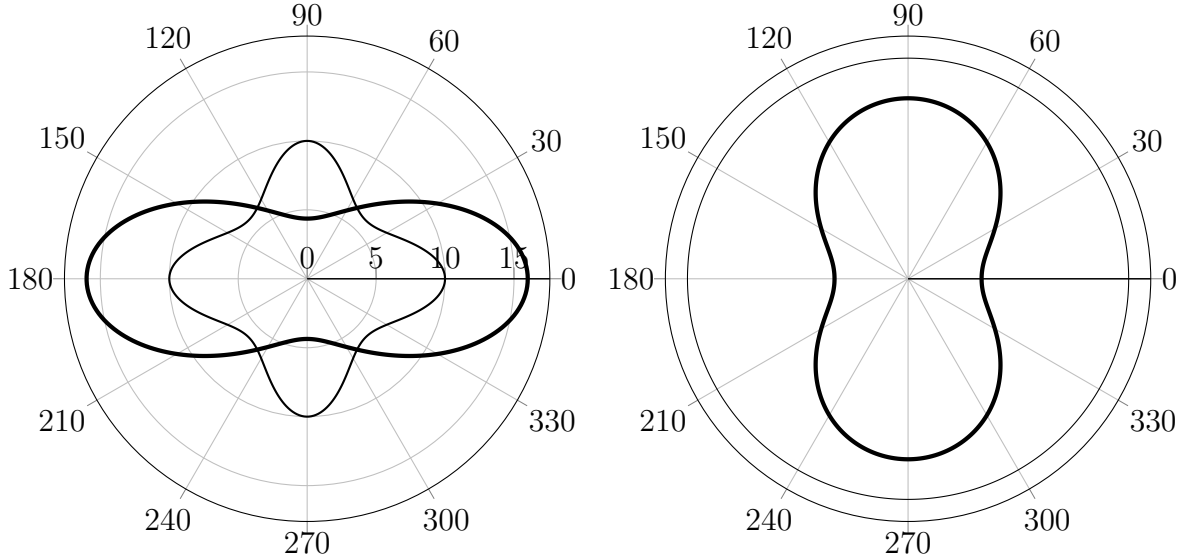
\begin{figure}[h]
\begin{center}
\begin{tikzpicture}
\begin{polaraxis} [width=0.5\textwidth]
 \addplot[ultra thick,domain=0:360,samples=360]
{0.001/(0.0001375+0.0000083333*cos(4*x)+0.0000833332*cos(180-2*x))};
\addplot[thick,domain=0:360,samples=360]
{0.001/(4*(0.0000333333-0.0000083333*cos(4*x)))};
\end{polaraxis}
\end{tikzpicture}
\begin{tikzpicture}
\begin{polaraxis} [width=0.5\textwidth,ytick={-0.5}]
 \addplot[ultra thick,domain=0:360,samples=360]
{1.5+(-0.0000708333+0.00000833333*cos(4*x))/(0.0001375+0.0000083333*cos(4*x)+0.0000833332*cos(180-2*x))};
\addplot[ domain=0:360,samples=360]{1.5};
\end{polaraxis}
\end{tikzpicture}
\caption{Polar diagrams for the layer of Example 5; left: $E_1(\theta),G_{12}(\theta)$, respectively thick and thin curves; right:  $\nu_{12}(\theta)$ (the thin circular line corresponds to zero: inside it, $\nud<0$).}
\label{fig:15}
\end{center}
\end{figure}

{\bf Example 5}: as an example of a material that can produce TAALs, let us consider a material with $E_1=16, E_2=4.364, G_{12}=10,\nu_{12}=-1\Rightarrow\Q_{11}=22, \Q_{22}=6,\Q_{12}=-6,\Q_{66}=20$, which gives $T_0=10,T_1=2,R_1=2,\Phi_1=0^\circ\Rightarrow\tau=0.2,\sigma=0.2$, so a point of set T, indicated by label "5" in Fig. \ref{fig:13}. The polar diagrams of $E_1(\theta),G_{12}(\theta)$ and $\nu(\theta)$ are shown in Fig. \ref{fig:15}. As seen above, the material must be itself totally auxetic, as will be any laminate fabricated with such a layer.

\subsubsection{Existence of classical anisotropic plies for  fabricating TAALs}
We ponder now, like in the case of $R_1=0$ materials, whether or not $R_0$-orthotropic plies able to realize TAALs can actually exist. A $R_0$-orthotropic ply can be obtained reinforcing an isotropic matrix with an equal volume fraction of fibres aligned in  two direction rotated of $45^\circ$, \cite{vannucci02joe}. The evaluation of the ply technical moduli can then be done using the classical laminated plates theory (CLPT), \cite{jones,gay14}, considering each $R_0$-orthotropic ply as the assemblage of two identical layers rotated of $45^\circ$. For each one of these two layers, the technical moduli are expressed as function of $E,\nu$, eq. (\ref{eq:adimparameters}), and of the volume fraction of the fibres, $\vf$, like in eq. (\ref{eq:romadim}). Then, the Cartesian components of the reduced stiffness tensor of each one of the two identical layers are calculated by eq. (\ref{eq:Qorthply}). The components $\Q_{ij}$ of the in-plane stiffness tensor  for the $R_0$-orthotropic ply can then be calculated using the CLPT and finally the technical moduli computed inverting the equations in  (\ref{eq:Qorthply}). Then, the dimensionless parameters $\alpha$ and $\eps$ for the $R_0$-orthotropic ply can  be calculated,  they are expressed as functions of the dimensionless parameters $E,\nu$ and $\vf$. These functions are omitted here because their expression is too much long. 

\begin{figure}
\centering
\includegraphics[width=0.4\textwidth]{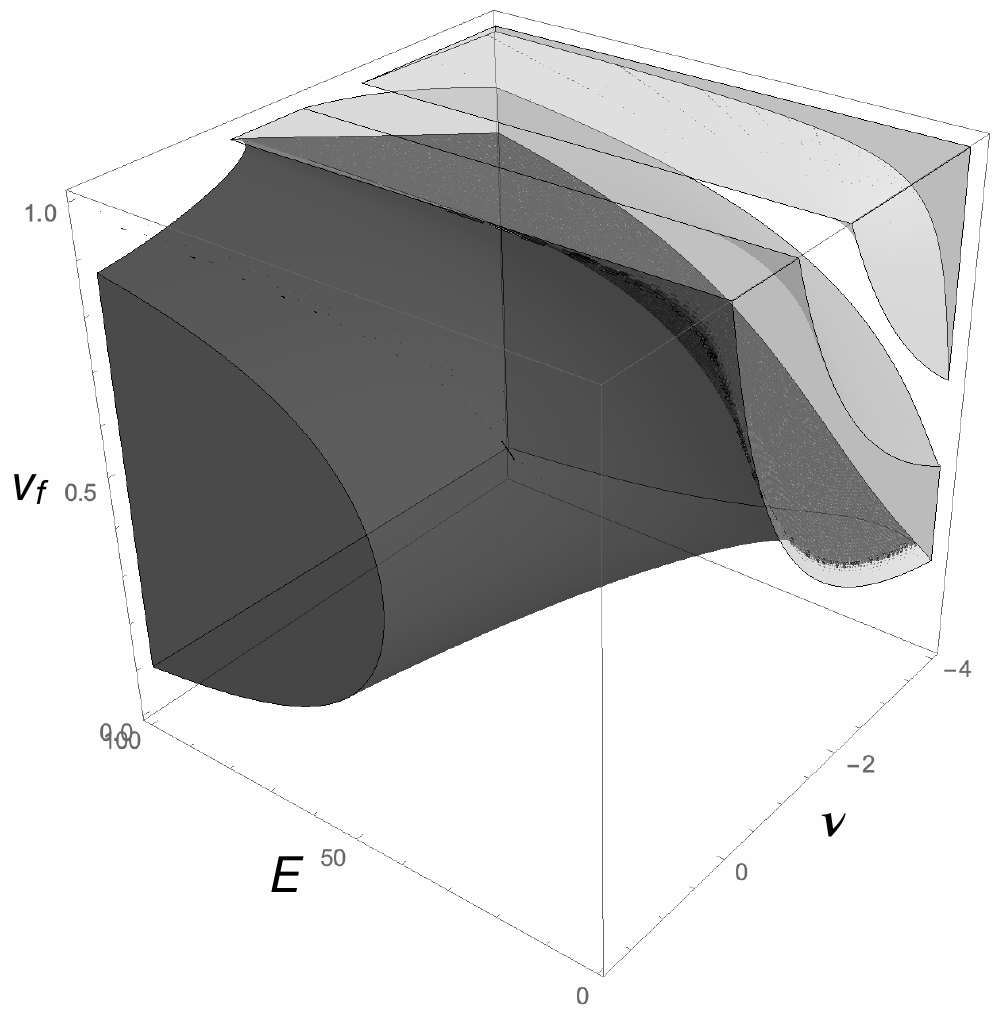}
\caption{Domain of existence of plies with $R_0=0$ suitable for the fabrication of TAALs or PAALs; the darker part corresponds to PAALs, the lighter one to TAALs. }
\label{fig:16}
\end{figure}

If now $\alpha$ and $\eps$ are inserted into eq. (\ref{eq:boundsmoduliadimR0ply}) or, alternatively, the moduli of the layer directly into eq. (\ref{eq:boundsmoduliR0ply}), the bounds for the existence of TAALs composed of $R_0$-orthotropic plies are found as functions of $E,\nu$ and $\vf$. Too complicate to be written here, the admissible domain in the space $(E,\nu,\vf)$ is anyway represented in Fig. \ref{fig:16}. It can be seen that, like in the previous case of $R_1=0$ plies, and also of unidirectional layers, \cite{vannucci24b}, to obtain TAALs layers with $\nu<0$ are needed. In other words, also with this materials it is impossible to obtain TAALs unless layers composed of either an auxetic matrix or of auxetic fibres are used: anisotropy  is not sufficient to obtain TAALs.

\subsection{Partially Auxetic Anisotropic Laminates (PAALs)}
\subsubsection{Polar conditions}
The auxeticity condition in this case is $\eta(\xi_3,\theta)<0$ for at least one $\theta$, hence, for $\theta=\pi/4$, we get
\be
\tilde{\eta}(\xi_3)=2\tau-4\sigma^2\xi_3^2-1<0,
\ee
which gives for $\xi_3\in[0,1]$
\be
\xi_3>\frac{\sqrt{2\tau-1}}{2\sigma},
\ee
a value that is real only for $\tau\geq1/2$, hence out of the TAAL zone, cf. eq. (\ref{eq:existtaalR0}). 
Because $\xi_3<1$, it must be
\be
\frac{\sqrt{2\tau-1}}{2\sigma}<1\Rightarrow\tau<\frac{1}{2}+2\sigma^2,
\ee
so finally, taking into account the elastic bound (\ref{eq:boundsR0adim}), the subset of the space $(\sigma,\tau)$ where PAALs exist is bounded by the conditions
\be
\label{eq:bornespolairesPAALs}
\max\left\{\frac{1}{2},2\sigma^2\right\}<\tau<\frac{1}{2}+2\sigma^2,
\ee
represented in Fig. \ref{fig:13} with the label P. Coming back to the polar parameters $T_0,T_1,R_1$ we get
\be
\max\left\{\frac{T_0}{2},2\frac{R_1^2}{T_0}\right\}<T_1<\frac{T_0}{2}+2\frac{R_1^2}{T_0}.
\ee

\subsubsection{Cartesian conditions}
The above bounds are easily transformed into Cartesian relations:
\be
\label{eq:cartcondR0PAALs}
\begin{array}{l}
\Q_{11}+6\Q_{12}+\Q_{22}-2\Q_{66}>0,\medskip\\
\Q_{11}^2-6\Q_{11}\Q_{22}+\Q_{22}^2+4\Q_{12}(\Q_{12}-\Q_{66})-2\Q_{66}(\Q_{11}+\Q_{22})<0,\medskip\\
3\Q_{11}^2+12\Q_{12}^2+3\Q_{22}^2+4\Q_{66}^2-2\Q_{11}(2\Q_{12}+5\Q_{22})-4\Q_{12}(\Q_{22}+4\Q_{66})>0.
\end{array}
\ee
Once more considering the definitions of $\alpha, \beta$ and $\eps$, eqs. (\ref{eq:alphabeta}), (\ref{eq:epsdef}), along with eq. (\ref{eq:epsalfabeta}), the above bounds resume to the dimensionless conditions
\be
\alpha>0,\ \ \ \beta>\frac{1}{2}(\alpha-1)^2,\ \ \ (\alpha-1)^2+(\beta-1)^2>1,
\ee
that determine the subset denoted by label P in Fig. \ref{fig:14} a). Coming back to the $\Q_{ij}$s, the last conditions are
\be
\Q_{12}>0,\ \ \ 2\Q_{22}\Q_{66}>(\Q_{12}-\Q_{22})^2,\ \ \ (\Q_{12}-\Q_{22})^2+(\Q_{66}-\Q_{22})^2>\Q_{22}^2,
\ee
much simpler than and equivalent to eq. (\ref{eq:cartcondR0PAALs}).

\subsubsection{Technical moduli conditions}
Proceeding like in the case of the TAALs, the following auxeticity conditions for PAALs in terms of technical moduli are get:
\be
\label{eq:condmoduliR0PAALs}
\begin{array}{l}
E_1^3-2E_1^2(3E_2+2G_{12})+4E_2^2G_{12}\nud^2(1+2\nud)+\smallskip\\
+E_1E_2[E_2-4G_{12}(1+2\nud)+4(E_2+G_{12})\nud^2]<0,\bigskip\\
\dfrac{E_1(E_1+E_2-4G_{12})+6E_1E_2\nud+4E_2G_{12}\nud^2}{E_1-E_2\nud^2}>0,\bigskip\\
3E_1^4+16E_2^2G_{12}^2\nud^4-2E_1^3E_2(5+2\nud)-32E_1E_2G_{12}\nud^2(G_{12}-\nud E_2)-\smallskip\\-E_1^2[32E_2G_{12}\nud-16G_{12}^2+E_2^2(4(1-3\nud)\nud-3)]>0.
\end{array}
\ee
Also in this case, the dimensionless form of the above bounds is particularly simple:
\be
\label{eq:boundsR0plyadimPAALs}
\alpha>0,\ \ \ \eps>\alpha^2,\ \ \ 8\alpha^2+(\eps-1)^2-4\alpha(1+\eps)>0.
\ee
These conditions delimit the subset of the space $(\alpha,\eps)$ denoted by label P in Fig. \ref{fig:14} b). Also in this case, going back to the moduli, we get conditions equivalent to those in eq. (\ref{eq:condmoduliR0PAALs}) but much simpler:
\be
\nud>0,\ \ \ E_1>\nud^2E_2,\ \ \ (E_1-E_2)^2-4\nud E_2(E_1+E_2)+8\nud^2E_2^2>0.
\ee

\subsubsection{Further considerations and  examples}
\begin{figure}[h]
\begin{center}
\begin{tikzpicture}
\begin{polaraxis} [width=0.5\textwidth]
 \addplot[ultra thick,domain=0:360,samples=360]
{0.001/(0.0000839285 + 0.0000142857*cos(4*x) + 0.0000714284*cos(180-2*x))};
\addplot[thick,domain=0:360,samples=360]
{0.001/(4*(0.0000392857 - 0.0000142857*cos(4*x)))};
\end{polaraxis}
\end{tikzpicture}
\begin{tikzpicture}
\begin{polaraxis} [width=0.5\textwidth,ytick={-0.5}]
 \addplot[ultra thick,domain=0:360,samples=360]
{1.+(-5.3571*10^(-6) + 0.0000142857*cos(4*x))/(0.0000839285 + 0.0000142857*cos(4*x) +0.0000714284*cos(180-2*x))};
\addplot[ domain=0:360,samples=360]{1.};
\end{polaraxis}
\end{tikzpicture}
\caption{Polar diagrams for the layer of Example 6; left: $E_1(\theta),G_{12}(\theta)$, respectively thick and thin curves; right:  $\nu_{12}(\theta)$ (the thin circular line corresponds to zero: inside it, $\nud<0$).}
\label{fig:17}
\end{center}
\end{figure}
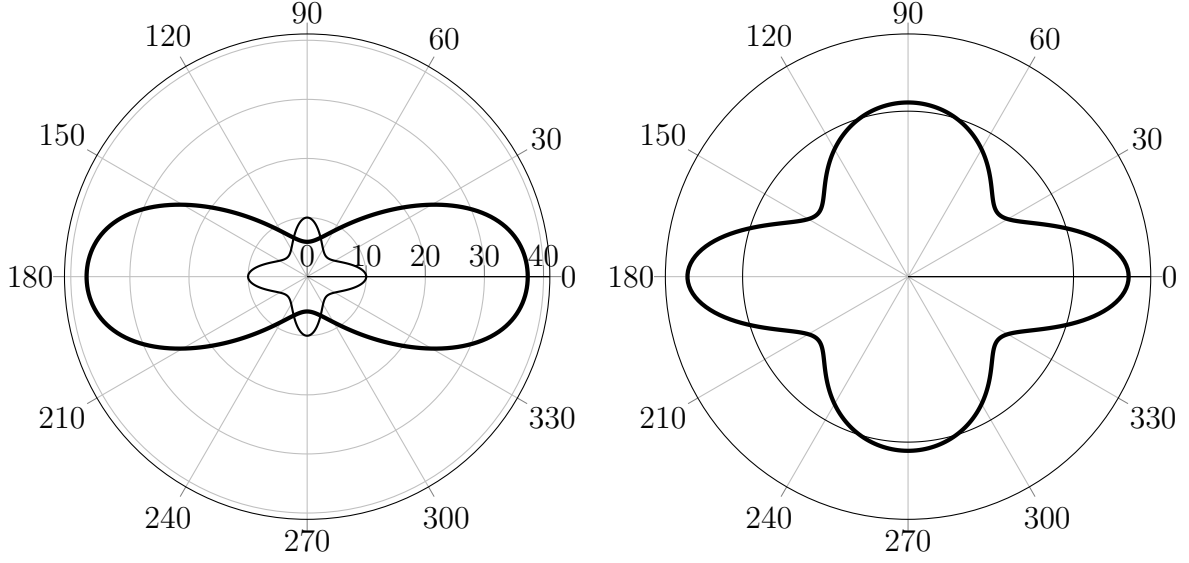

{\bf Example 6}: as an example of a material that can produce PAALs, let us consider a material with $E_1=37.333, E_2=5.895, G_{12}=10,\nu_{12}=0.333\Rightarrow\Q_{11}=38, \Q_{22}=6,\Q_{12}=2,\Q_{66}=20$, which gives $T_0=10,T_1=6,R_1=4,\Phi_1=0^\circ\Rightarrow\tau=0.6,\sigma=0.4$, so a point of set T, indicated by label "6" in Fig. \ref{fig:13}. The polar diagrams of $E_1(\theta),G_{12}(\theta)$ and $\nu(\theta)$ are shown in Fig. \ref{fig:17}.  
It can be seen that the layer is itself partially auxetic. 

Like in the case of materials with $R_1=0$, the question is: is it possible to fabricate a PAAL with $R_0=0$ layers having $\nud(\theta)>0\ \forall\theta$? To this end, it is necessary to have (in the case of a single layer, $\xi_3=1$)
\be
\min_\theta\eta(\xi_3=1,\theta)=2\tau-2\sigma^2(1-\cos4\theta)-1>0;
\ee
 the minimum is clearly get for $\theta=\pi/4$, which gives the condition
 \be
 \eta\left(1,\frac{\pi}{4}\right)=2\tau-4\sigma^2-1>0\ \iff\ \tau>\frac{1}{2}+2\sigma^2.
 \ee
Because this condition is incompatible with the domain defining PAALs, eq. (\ref{eq:bornespolairesPAALs}), it is impossible to fabricate PAALs with layers that are not partially auxetic themselves.

Also for these materials it is possible to find the analytical expression of the minimum of $\nud^\A(\theta)$; proceeding like in the case of $R_1=0$ layers, it can be seen that 
\be
{\nud^\A}_{\min}=1-\frac{1}{\sqrt{2\tau-4\sigma^2}} \ \mathrm{for}\ \xi_3=1,\ \theta=\frac{1}{2}\arccos\frac{1-\sqrt{2\tau-4\sigma^2}}{2\sigma}.
\ee
Also in this case, hence, it is impossible to obtain a PAAL done with  partially auxetic $R_0=0$ plies having  a  minimum of the Poisson's ratio lower than that of the single layer.

A consequence of this result, is that the only  design problem for PAALs made of $R_0$-orthotropic plies  is, possibly, still the maximization of the auxetic zone. Proceeding like in the case of $R_1=0$ materials, from eq. (\ref{eq:auxeticR0cond}) we get that the condition maximizing the auxetic zone is 
\be
\max_{\xi_3\in[0,1]}\psi(\xi_3):=\frac{1-2\tau+2\sigma^2\xi_3^2}{2\sigma^2\xi_3^2}.
\ee
It is immediately recognized that the maximum of $\psi(\xi_3)$ is get for $\xi_3=1$:  the largest auxetic zone $\Delta\alpha$ is that of the single layer:
\be
\Delta\alpha=\frac{\pi}{2}-\frac{1}{2}\arccos\frac{1-2\tau+2\sigma^2}{2\sigma^2}.
\ee
So, also for this problem any PAAL composed of $R_0=0$ layers will have an auxetic zone smaller than that of the single ply.

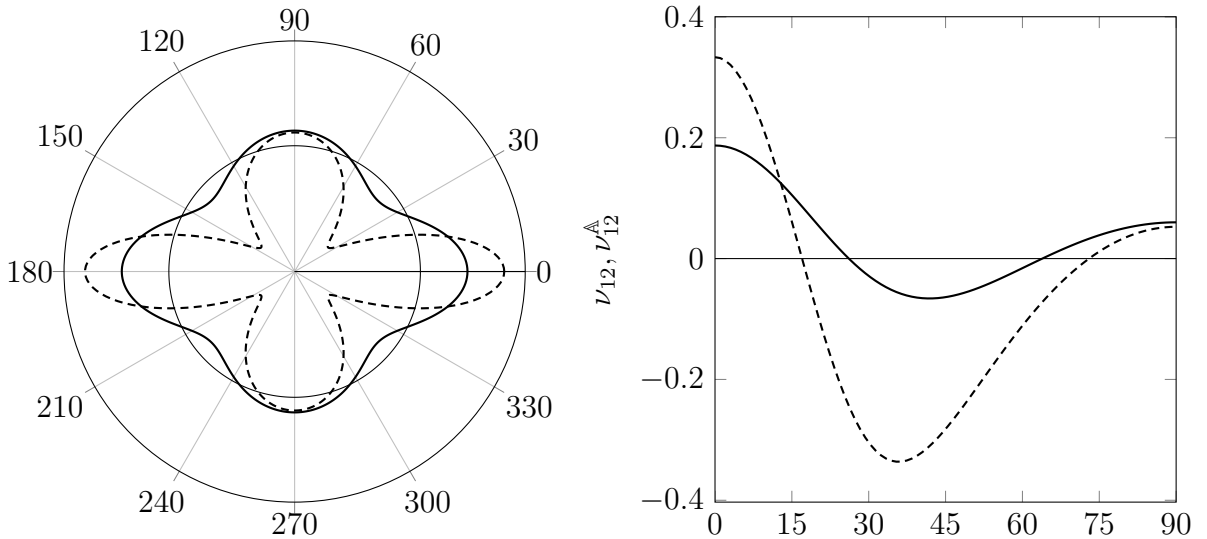
\begin{figure}[h]
\begin{center}
\begin{tikzpicture}
\begin{polaraxis} [width=0.48\textwidth,ytick={-0.5}]
 \addplot[ thick,domain=0:360,samples=360]
{0.5+(1.1365*10^(-6) + 4.54545*10^(-6)*cos(4*x))/(0.0000579545+4.54545*10^(-6)*cos(4*x) + 
   0.0000321412*cos(180-2*x))};
\addplot[ thick,densely dashed,domain=0:360,samples=360]
{0.5+(-5.3571*10^(-6) + 0.0000142857*cos(4*x))/(0.0000839285 + 0.0000142857*cos(4*x) +0.0000714284*cos(180-2*x))};
\addplot[ domain=0:360,samples=360]{0.5};
\end{polaraxis}
\end{tikzpicture}
\begin{tikzpicture}
\begin{axis} [width=0.48\textwidth,height=0.5\textwidth,xtick={0,15,30,45,60,75,90},xmin=0,xmax=90,ylabel=${\nud,\nud^\A}$]
 \addplot[ thick,domain=0:90,samples=90]
{(1.1365*10^(-6) + 4.54545*10^(-6)*cos(4*x))/(0.0000579545+4.54545*10^(-6)*cos(4*x) + 
   0.0000321412*cos(180-2*x))};
\addplot[ thick,densely dashed,domain=0:90,samples=90]
{(-5.3571*10^(-6) + 0.0000142857*cos(4*x))/(0.0000839285 + 0.0000142857*cos(4*x) +0.0000714284*cos(180-2*x))};
\addplot[domain=0:90]{0};
\end{axis}
\end{tikzpicture}
\caption{Polar and Cartesian diagrams of $\nud(\theta)$, dashed line, and of $\nud^\A(\theta)$, solid line, for the  material of Example 6 and the laminate  with $\xi_1=0.707$ made with this material (the thin circular line corresponds to zero: inside it, $\nud<0$). }
\label{fig:18}
\end{center}
\end{figure}

{\bf Example 7}: let us consider the case of a laminate composed of the material in Example 6 and having $\xi_3=0.707$. It is the case, for instance, of an angle-ply laminate with  ply angles 
\be
\delta=\pm\frac{1}{2}\arccos\xi_3=\pm22.5^\circ.
\ee
For the single layer, it is ${\nud}_{min}=-0.336$, at the angle $\theta=35.8^\circ$ and $\Delta\alpha=56^\circ$, while for the laminate ${\nud^\A}_{min}=-0.066$ for $\theta=41.8^\circ$ and $\Delta\alpha=37.8^\circ$. The diagrams of $\nud(\theta)$ and $\nud^\A(\theta)$ are shown in Fig. \ref{fig:18}. It can be seen that the single layer has a lower minimum of the Poisson's ratio and a wider auxetic zone than the laminate, according to the above results. 

\subsubsection{Existence of classical anisotropic plies for  fabricating PAALs}
Just like for TAALs, we ponder now whether or not $R_0$-orthotropic plies able to realize PAALs can actually exist. The procedure is exactly the same sketched above  for the TAALs, only the bounds change, now they are represented by eq. (\ref{eq:boundsR0plyadimPAALs}). The domain defined by these bounds is represented in Fig. \ref{fig:16} and it can be seen that such a domain exists also for $\nu>0$: $R_0$-orthotropic plies composed of non auxetic matrix and fibres and able to realize PAALs can really exist. Like in the case of unidirectional layers, but unlike $R_1=0$ plies, anisotropy is sufficient to obtain auxetic  laminates.

\section{The strage case of $r_0-$orthotropic materials}
The special anisotropy of paper was discovered experimentally in 1951 by Horio and Onogi, \cite{horio51}. The rather unusual fact highlighted by them, was the isotropy of the shear modulus $G_{12}$, that, along with the other peculiarity of an angular variation of the Young's modulus and of the Poisson's ratio typical of quantities of a second-rank tensor, gave this material very particular characteristics that did not seem to be attributable to the classical theory of anisotropic elasticity. The unusual elasticity of the paper has been the subject of different studies, see e.g. \cite{campbell61,ostoja2000}, and in \cite{vannucci10joe} it has been shown that actually this peculiar behavior corresponds to $r_0-$orthotropy, i.e. to the special case where the compliance tensor $\Sy$ does not depend upon the harmonic varying with $4\theta$ but just on that changing with $2\theta$, like second-order tensors. 

\subsection{Polar conditions for the auxeticity of $r_0-$orthotropic plies}
Because this special case of orthotropy is determined by a condition on $\Sy$, it is worth, and easier, to state directly the auxeticity condition using its polar parameters. Actually, cf. \cite{vannucci_libro}, if $r_0=0$, then
\be
\nud(\theta)=\frac{t_0-2t_1}{t_0+2t_1+4r_1\cos2\theta}.
\ee
Because the denominator is the polar expression of $\Sy_{11}(\theta)$, which is positive for any direction $\theta$, then the auxeticity condition is simply
\be
t_0<2t_1,
\ee
i.e. for $r_0=0$ materials, auxeticity is determined by an isotropic condition, although the Poisson's ratio remains an anisotropic parameter, varying like $2\theta$. This implies that  such materials are either totally auxetic, i.e. $\nud(\theta)<0\ \forall\theta$, or totally non-auxetic, i.e. $\nud(\theta)>0\ \forall\theta$, partial auxeticity is excluded.

Using eq. (\ref{eq:componentipolarisouplesse}) the above auxeticity condition can also be rewritten using the polar parameters of the reduced stiffness tensor $\Q$:
\be
T_0T_1+R_1^2>2T_1^2.
\ee
Because for these materials it is
\be
\label{eq:auxr01}
R_0=\frac{R_1^2}{T_1},\ \ \ K=0\Rightarrow\Phi_0=\Phi_1,
\ee
then we get the inequality
\be
T_1<\frac{T_0+R_0}{2},
\ee
and finally, still introducing the dimensionless parameters $\tau$ and $\rho$,
\be
\label{eq:auxetiquer0adim}
\tau<\frac{1+\rho}{2}.
\ee

\subsection{Polar conditions for the auxeticity of $r_0-$orthotropic laminates}
Applying eq. (\ref{eq:auxr01}) to the auxeticity condition (\ref{eq:aux1}) for a laminate and choosing $\Phi_0^\A=\Phi_1^\A=0$ to fix the frame  gives
\be
2T_1(T_0-R_0\xi_3^2)-T_0^2+R_0^2\xi_1^2+2T_1R_0(\xi_3^2-\xi_1)\cos4\theta<0
\ee
and if the dimensionless parameters $\tau$ and $\rho$ are introduced again
\be
\label{eq:auxeticitylaminatesr0A}
2\tau(1-\rho\xi_3^2)-1+\rho^2\xi_1^2+2\tau\rho(\xi_3^2-\xi_1)\cos4\theta<0.
\ee
The complete solution to the problem hold by this condition is rather difficult to be found. However, a situation is rather intriguing: is it possible to have an auxetic laminate composed by $r_0$-orthotropic plies with $r_0^\A=0$, i.e. $r_0-$orthotropic itself? In fact, unlike for $R_0$ and $R_1$, that are stiffness moduli, $r_0$, a compliance parameter, is not preserved by the homogenization giving the elastic properties of the laminate. 

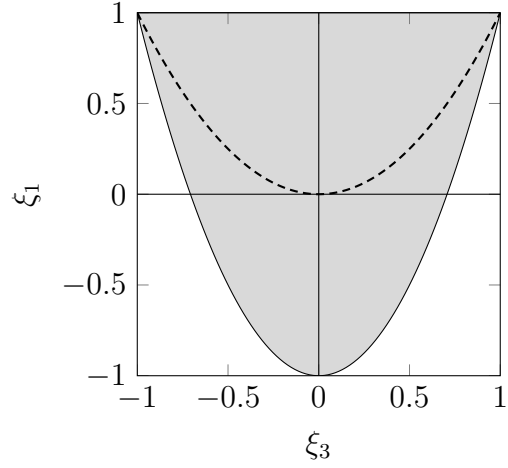
\begin{figure}
\begin{center}
\begin{tikzpicture}
\begin{axis} [xmin=-1,xmax=1.,xlabel=$\xi_3$,
 ymin=-1,ymax=1,ylabel=$\xi_1$, width=0.3\textwidth,height=0.3\textwidth,scale only axis,samples=100]
\addplot[ name path=A,domain=-1:1]{2*x^2-1};
\addplot[ name path=B,domain=-1:1]{1};
\addplot[domain=-1:1]{0};
\addplot[domain=-1:1]coordinates{(0,-1)(0,1)};
\addplot[ thick,densely dashed,name path=C,domain=-1:1]{x^2};
\addplot[gray,fill opacity=0.3] fill between[of=A  and B] ;
\end{axis}
\end{tikzpicture}
\caption{Domain of admissible lamination parameters for orthotropic laminates. The dashed curve is the parabola (\ref{eq:parabola}).}
\label{fig:19}
\end{center}
\end{figure}

Eq. (\ref{eq:auxr01}) applies not only to $\Q$, but  to any $r_0-$orthotropic tensor, hence to $\A$ too:  
\be
r_0^\A=0\iff R_0^\A=\frac{{R_1^\A}^2}{T_1^\A}.
\ee 
Then, because $T_1^\A=T_1$ and by eqs. (\ref{eq:R0laminato}) and (\ref{eq:R1laminato}), we get
\be
R_0\xi_1=\frac{R_1^2\xi_3^2}{T_1}
\ee
and finally, because  plies are $r_0-$orthotropic, by eq. (\ref{eq:auxr01})
\be
\label{eq:parabola}
\xi_1=\xi_3^2.
\ee
This is a constraint on the stack: laminates observing the above inequality and composed by $r_0-$orthotropic plies will have $r_0^\A=0$. Such condition is perfectly possible: Miki, \cite{Miki82}, has shown that the set of admissible lamination parameters is the set of the plane $(\xi_3,\xi_1)$ bounded by the conditions
\be
2\xi_3^2-1\leq\xi_1\leq1,
\ee
shown in Fig. \ref{fig:19}. The parabola (\ref{eq:parabola}) is clearly inside it.

A question remains: are laminates so obtained auxetic? If eq. (\ref{eq:parabola}) is injected into the auxeticity condition (\ref{eq:auxeticitylaminatesr0A}), this last becomes
\be
\rho^2\xi_3^4-2\tau\rho\xi_3^2+2\tau-1<0,
\ee
whose solution, for $\xi_3\in[0,1]$ is
\be
\label{eq:auxeticitylaminator0Aconr0}
\sqrt{\frac{2\tau-1}{\rho}}<\xi_3<1.
\ee
This solution is admissible if
\be
\sqrt{\frac{2\tau-1}{\rho}}<1,
\ee
i.e. if eq. (\ref{eq:auxetiquer0adim}) is satisfied. So, if the $r_0-$orthotropic ply is auxetic, then  conditions (\ref{eq:parabola}) and (\ref{eq:auxeticitylaminator0Aconr0}) ensures that the laminate will be (totally) auxetic and $r_0-$orthotropic.

\subsection{Further considerations and examples}
We end this part showing some examples of auxetic plies and laminates with $r_0=0$.

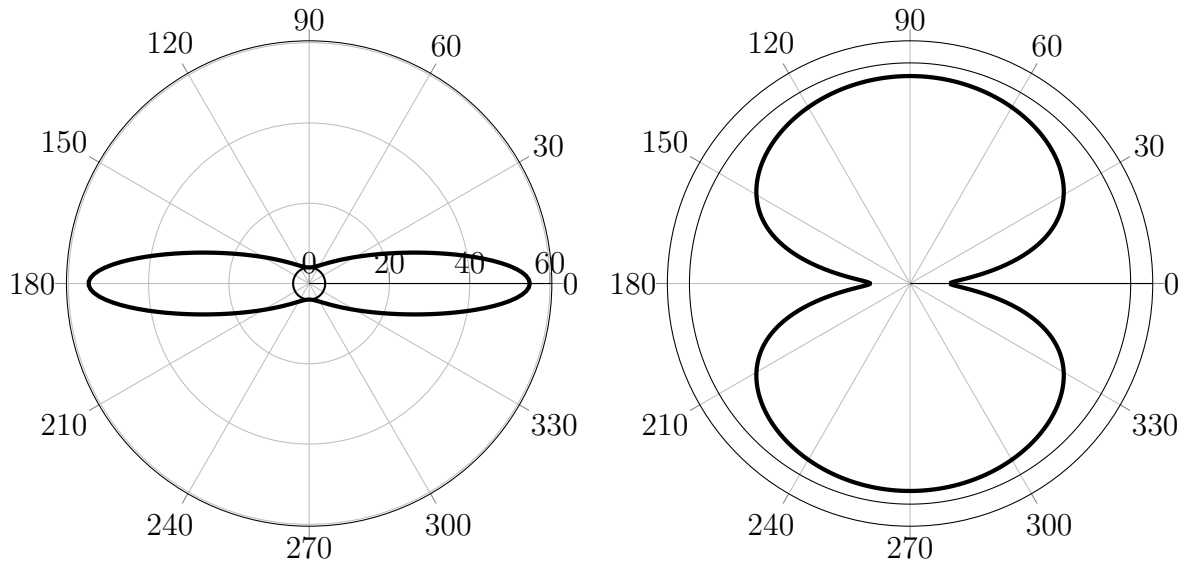
\begin{figure}[h]
\begin{center}
\begin{tikzpicture}
\begin{polaraxis} [width=0.5\textwidth]
 \addplot[ultra thick,domain=0:360,samples=360]
{0.001/(0.000133953 + 0.00011576*cos(180 - 2*x))};
\addplot[thick,domain=0:360,samples=360]{4};
\end{polaraxis}
\end{tikzpicture}
\begin{tikzpicture}
\begin{polaraxis} [width=0.5\textwidth,ytick={-0.5}]
 \addplot[ultra thick,domain=0:360,samples=360]
{0.6-(8.9383*10^(-6)/(0.000133953 + 0.00011576*cos(180-2*x))};
\addplot[ domain=0:360,samples=360]{0.6};
\end{polaraxis}
\end{tikzpicture}
\caption{Polar diagrams for the layer of Example 7; left: $E_1(\theta),G_{12}(\theta)$, respectively thick and thin curves; right:  $\nu_{12}(\theta)$ (the thin circular line corresponds to zero: inside it, $\nud<0$). To remark the isotropy of the shear modulus $G_{12}$.}
\label{fig:20}
\end{center}
\end{figure}

{\bf Example 7}: let us take as an example of $r_0-$orthotropic layer that of a ply with $E_1=54.943, E_2=4.004, G_{12}=4,\nu_{12}=-0.491\Rightarrow\Q_{11}=55.924, \Q_{22}=4.076,\Q_{12}=-2,\Q_{66}=8$, which gives $T_0=10,T_1=7,R_0=6,R_1=6.481,\Phi_0=\Phi_1=0^\circ\Rightarrow\tau=0.7,\rho=0.6,\sigma=0.648$, indicated by label "7" in both Figs. \ref{fig:1} and  \ref{fig:13}. The polar diagrams of $E_1(\theta),G_{12}(\theta)$ and $\nu(\theta)$ are shown in Fig. \ref{fig:20}.  

\begin{figure}[h]
\begin{center}
\begin{tikzpicture}
\begin{polaraxis} [width=0.48\textwidth,ytick={-0.5}]
 \addplot[ thick,domain=0:360,samples=360]
{0.6-(1.35177/(162.801 - 103.696*cos(2*x)};
\addplot[ thick,densely dashed,domain=0:360,samples=360]
{0.6-(8.9383*10^(-6)/(0.000133953 + 0.00011576*cos(180-2*x))};
\addplot[ domain=0:360,samples=360]{0.6};
\end{polaraxis}
\end{tikzpicture}
\begin{tikzpicture}
\begin{axis} [width=0.48\textwidth,height=0.5\textwidth,xtick={0,15,30,45,60,75,90},xmin=0,xmax=90,ylabel=${\nud,\nud^\A}$]
 \addplot[ thick,domain=0:90,samples=90]
{-(1.35177/(162.801 - 103.696*cos(2*x)};
\addplot[ thick,densely dashed,domain=0:90,samples=90]
{-(8.9383*10^(-6)/(0.000133953 + 0.00011576*cos(180-2*x))};
\addplot[domain=0:90]{0};
\end{axis}
\end{tikzpicture}
\caption{Polar and Cartesian diagrams of $\nud(\theta)$, dashed line, and of $\nud^\A(\theta)$, solid line, for the  material of Example 7 and the laminate  with $\xi_1=0.7056$ and $\xi_3=0.84$ made with this material (the thin circular line corresponds to zero: inside it, $\nud<0$). }
\label{fig:21}
\end{center}
\end{figure}
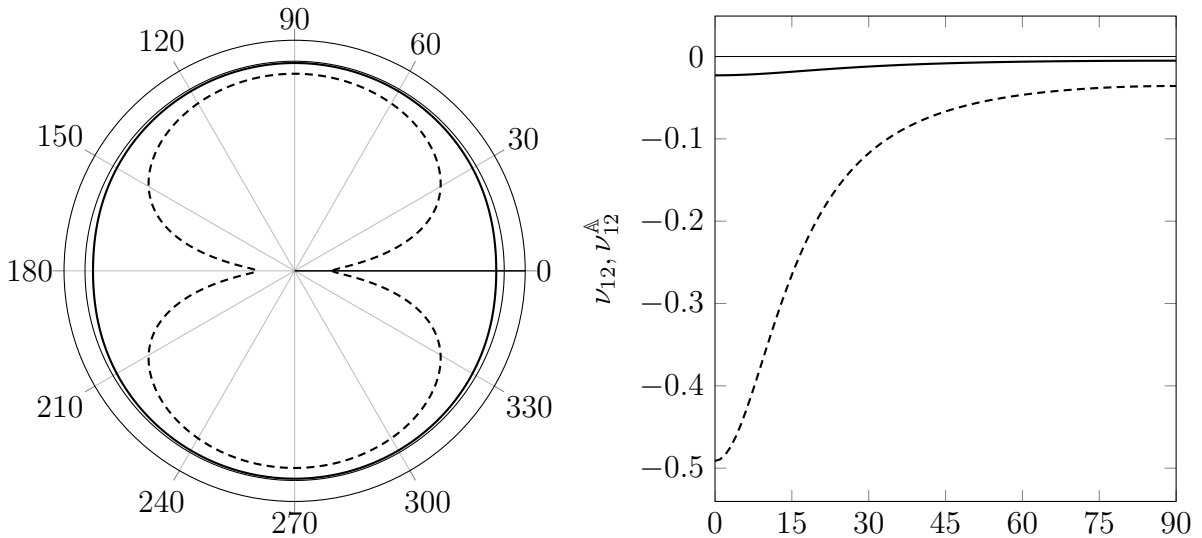

As an example of auxetic laminate composed of plies of this material, let us consider the case of a laminate with $\xi_1=0.7056,\xi_3=0.84$, so that conditions (\ref{eq:parabola}) and (\ref{eq:auxeticitylaminator0Aconr0}) are satisfied. The laminate so obtained is totally auxetic, see Fig. \ref{fig:21}, but the minimum of the Poisson's ratio is higher than that of the layer.
As a final special case of auxetic laminate obtained by such a material, let us consider the one with $\xi_1=\xi_3=0$. As known, this corresponds to an isotropic laminate and indeed it is also the minimum of the parabola (\ref{eq:parabola}). Such a laminate can be obtained, e.g., using again the Werren and Norris rules, \cite{werren53}, like for example the same 6-ply symmetric laminate with orientations $[0^\circ,60^\circ,-60^\circ]_{sym}$ already used above. In that case, the laminate will have a Young's modulus $E_1^\A=23.333$, a shear modulus $G_{12}^\A=10$ and a Poisson's ratio $\nud^\A=0.166\ \forall\theta$. The laminate is hence non auxetic at all, because condition (\ref{eq:auxeticitylaminator0Aconr0}) is not satisfied. This is a rather strange case of a laminate which, although composed of totally auxetic plies, is totally non auxetic.

\section{Final considerations and conclusion}
The subject of this paper was a purely theoretical one: to explore the possible auxeticity of laminates made of specially orthotropic plies. Namely, the three possible, and rather different, cases of $R_1-, R_0$ and $r_0-$orthotropy have been considered, looking for the conditions needed to obtain totally or partially auxetic laminates.
The results are rather surprisingly, because clear differences among the three cases and with the case of unidirectional plies, previously analyzed, emerge:
\begin{itemize}
\item plies having $R_1=0$, a very common case in practical applications, can be auxetic and produce auxetic laminates only if they are composed by one of the two phases, either the matrix or the fibres, that is itself auxetic. For this kind of materials, anisotropy is not sufficient by itself to produce an auxetic, although partial, behavior, ans this is a major difference with respect to all the other cases; seemingly, this derives from the fact that the anisotropic behavior of this kind of plies is not sufficient to produce auxeticity, algebraically, with no doubts, it is due to the absence of the harmonics varying with $2\theta$ in the polar representation of the Cartesian components of the reduced stiffness tensor $\Q$;
\item plies having $R_0=0$, a very special case of anisotropy and not encountered yet in practical applications, can actually produce PAALs, but not TAALs, also if composed by classical, non auxetic phases, but only if the ply itself is partially auxetic, another difference with respect to the ordinarily orthotropic (unidirectional) plies;
\item the case of materials with $r_0=0$, like common paper, is very strange, for multiple reasons: first of all, a ply of this type can be only totally auxetic, not partially, although the Poisson's ratio remains variable with the orientation, but just like a quantity belonging to a second-rank tensor, a circumstance already known; then, it is possible, in theory, to produce TAALs with these plies, but it i also possible to fabricate laminates that are totally non auxetic, and namely this is always the case for an isotropic laminate.
\end{itemize}
A last remark concerns the mathematical method used to investigate this matter: the results show clearly that the polar formalism is very helpful in treating this problem and if used in a dimensionless form, it is possible to have simple graphical representations. The polar method helps very much also in looking for conditions stated using the Cartesian components of $\Q$ or the technical moduli, and also in this case the introduction of dimensionless parameters is of a great help in stating simple conditions. To find the same results using directly a Cartesian representation of anisotropic elasticity would have been much more cumbersome.

Finally, this paper is an exploration of a still hidden part of anisotropic elasticity, as such it has not yet had, for the while, a direct practical approach. Anyway, applications now could, hopefully, have a benefit from these results.

  \bibliographystyle{vancouver} 
 \bibliography{biblio}

\begin{thebibliography}{10}

\bibitem{Love}
Love AEH.
\newblock A treatise on the mathematical theory of elasticity.
\newblock New York, NY: Dover; 1944.

\bibitem{sokolnikoff}
Sokolnikoff IS.
\newblock Mathematical theory of elasticity.
\newblock New York, NY: McGraw-Hill; 1946.

\bibitem{cho2019}
Cho H, Seo D, Kim DN.
\newblock Mechanics of auxetic materials.
\newblock In: Schmauder S, Chen CS, editors. Handbook of mechanics of
  materials. Singapore: Springer; 2019. p. 733--757.

\bibitem{Almgren85}
Almgren RF.
\newblock {An isotropic three-dimensional structure with Poisson's ratio= -1}.
\newblock Journal of Elasticity. 1985;15:427--430.

\bibitem{Evans91}
Evans KE.
\newblock Auxetic polymers: a new range of materials.
\newblock Endeavour - New Series. 1991;15(4):170--174.

\bibitem{Evans_Nature91}
Evans KE, Nkansah MA, Hutchinson IJ, Rogers SC.
\newblock Molecular network design.
\newblock Nature. 1991;353:124.

\bibitem{Lakes1991}
Lakes R.
\newblock {Deformation mechanisms in negative Poisson's ratio materials:
  structural aspects}.
\newblock Journal of Materials Sciences. 1991;26:2287--2292.

\bibitem{Lakes1993}
Lakes R.
\newblock {Advances in negative Poisson's ratio materials}.
\newblock Advanced Materials. 1993;5(4):293--296.

\bibitem{Lakes2002}
Lakes R, Witt R.
\newblock {Making and characterizing negative Poisson's ratio materials}.
\newblock International Journal of Mechanical Engineering Education.
  2002;30(1):50--58.

\bibitem{Lakes2017}
Lakes R.
\newblock {Negative Poisson's-ratio materials: Auxetic solids}.
\newblock Annaul Review of Materials Research. 2017;47:63--81.

\bibitem{Milton92}
Milton GW.
\newblock {Composite materials with Poisson's ratios close to -1}.
\newblock Journal of the Mechanics and Physics of Solids.
  1992;40(5):1105--1137.

\bibitem{Prawoto12}
Prawoto Y.
\newblock {Seeing auxetic materials from the mechanics point of view: A
  structural review on the negative Poisson's ratio}.
\newblock Computational Materials Science. 2012;58:140--153.

\bibitem{Shukla22}
Shukla S, Behera BK.
\newblock {Auxetic fibrous structures and their composites: A review}.
\newblock Composite Structures. 2022;290(115530).

\bibitem{vannucci_libro}
Vannucci P.
\newblock Anisotropic elasticity.
\newblock Berlin, Germany: Springer; 2018.

\bibitem{Lekhnitskii}
Lekhnitskii SG.
\newblock Theory of elasticity of an anisotropic elastic body.
\newblock San Francisco, CA: English translation (1963) by {P. Fern},
  Holden-Day; 1950.

\bibitem{Herakovich84}
Herakovich CT.
\newblock {Composite laminates with negative through-the-thickness Poisson's
  ratios}.
\newblock Journal of Composite Materials. 1984;18(5):447--455.

\bibitem{Miki89}
Miki M, Murotsu Y.
\newblock {The peculiar behavior of the Poisson's ratio of laminated fibrous
  composites}.
\newblock JSME International Journal. 1989;32(1):67--72.

\bibitem{Clarke94}
Clarke JJ, Duckett RA, Hine PJ, Hutchinson IJ, Ward IM.
\newblock {Negative Poisson's ratios in angle-ply laminates: theory and
  experiment}.
\newblock Composites. 1994;25(9):863--868.

\bibitem{Hine97}
Hine PJ, Duckett RA, Ward IM.
\newblock {Negative Poisson's ratios in angles-ply laminates}.
\newblock Journal of Materials Science Letters. 1997;16:541--544.

\bibitem{Zhang98}
Zhang R, Yeh HL, Yeh HY.
\newblock {A preliminary study of negative Poisson's ratio of laminated fiber
  reinforced composites}.
\newblock Journal of Reinforced Plastics and Composites.
  1998;17(18):1651--1664.

\bibitem{Zhang99}
Zhang R, Yeh HL, Yeh HY.
\newblock {A discussion of negative Poisson's ratio design for composites}.
\newblock Journal of Reinforced Plastics and Composites.
  1999;18(17):1546--1556.

\bibitem{alderson2005}
Alderson KL, Simkins VR, Coenen VL, Davies PJ, Alderson A, Evans KE.
\newblock How to make auxetic fibre reinforced composites.
\newblock Physica Status Solidi. 2005;242(3):509--518.

\bibitem{Peel07}
Peel LD.
\newblock {Exploration of high and negative Poisson's ratio elastomer-matrix
  laminates}.
\newblock Physica Status Solidi. 2007;244(3):988--1003.

\bibitem{Shokrieh11}
Shokrieh MM, Assadi A.
\newblock {Determination of maximum negative Poisson's ratio for laminated
  fiber composites}.
\newblock Physica Status Solidi. 2011;248(5):1237--1241.

\bibitem{Veloso23}
Veloso C, Mota C, Cunha F, Sousa J, Fangueiro R.
\newblock A comprehensiv review on in-plane and through-the-thickness
  auxeticity in composite laminates for structural applications.
\newblock Journal of Composite Materials. 2023;57(26):4215--4223.

\bibitem{vannucci24b}
Vannucci P.
\newblock Anisotropic auxetic composite laminates: a polar approach.
\newblock Journal of Composite Materials. 2024;58(17):1947--1953.
\newblock Available from: \url{https://doi.org/10.1177/00219983241256335}.

\bibitem{vannucci24d}
Vannucci P.
\newblock Two optimization problems for the auxeticity properties of
  anisotropic laminates.
\newblock Journal of Composite Materials. 2024;Available from:
  \url{https://doi.org/10.1177/00219983241308084}.

\bibitem{Verchery79}
Verchery G.
\newblock Les invariants des tenseurs d'ordre 4 du type de
  l'{\'e}lasticit{\'e}.
\newblock In: {Proc. of Colloque Euromech 115 (Villard-de-Lans, 1979):
  Comportement m{\'e}canique des mat{\'e}riaux anisotropes}. Paris: Editions du
  CNRS; 1982. p. 93--104.

\bibitem{Meccanica05}
Vannucci P.
\newblock Plane anisotropy by the polar method.
\newblock Meccanica. 2005;40:437--454.

\bibitem{vannucci02joe}
Vannucci P.
\newblock A special planar orthotropic material.
\newblock Journal of Elasticity. 2002;67:81--96.

\bibitem{vannucci10joe}
Vannucci P.
\newblock On special orthotropy of paper.
\newblock Journal of Elasticity. 2010;99:75--83.

\bibitem{jones}
Jones RM.
\newblock Mechanics of composite materials. Second Edition.
\newblock Philadelphia, PA: Taylor \& Francis; 1999.

\bibitem{gay14}
Gay D.
\newblock {Composite Materials Design and Applications - Third Edition}.
\newblock Boca Raton, FL: CRC Press; 2014.

\bibitem{vannucci01joe}
Vannucci P.
\newblock On bending-tension coupling of laminates.
\newblock Journal of Elasticity. 2001;64:13--28.

\bibitem{vannucci23a}
Vannucci P.
\newblock On the mechanical and mathematical properties of the stiffness and
  compliance coupling tensors of composite anisotropic laminates.
\newblock Journal of Composite Materials. 2023;57(26):4197--4214.

\bibitem{vannucci23b}
Vannucci P.
\newblock On the thermoelastic coupling of anisotropic laminates.
\newblock Archive of Applied Mechanics. 2024;94:1121--1149.

\bibitem{vannucci01ijss}
Vannucci P, Verchery G.
\newblock Stiffness design of laminates using the polar method.
\newblock International Journal of Solids and Structures. 2001;38:9281--9294.

\bibitem{vannucci01cst}
Vannucci P, Verchery G.
\newblock A special class of uncoupled and quasi-homogeneous laminates.
\newblock Composites Science and Technology. 2001;61:1465--1473.

\bibitem{toponogov}
Toponogov VA.
\newblock {Differential geometry of curves and surfaces - A concise guide}.
\newblock Birkh{\"a}user; 2006.

\bibitem{vannucci_alg}
Vannucci P.
\newblock {Tensor algebra and analysis for engineers - With applications to
  differential geometry of curves and surfaces}.
\newblock Singapore: World Scientific; 2023.

\bibitem{TsaiHahn}
Tsai SW, Hahn T.
\newblock Introduction to composite materials.
\newblock Stamford, CT: Technomic; 1980.

\bibitem{ting}
Ting TCT.
\newblock Anisotropic elasticity.
\newblock Oxford, UK: Oxford University Press; 1996.

\bibitem{kelvin}
{Thomson - Lord Kelvin} W.
\newblock Elements of a mathematical theory of elasticity.
\newblock Philosophical Transations of the Royal Society. 1856;146:481--498.

\bibitem{kelvin1}
{Thomson - Lord Kelvin} W.
\newblock Mathematical theory of elasticity.
\newblock Encyclopedia Britannica. 1878;7:819--825.

\bibitem{vannucci24a}
Vannucci P.
\newblock {Complete set of bounds for the technical moduli in 3D anisotropic
  elasticity}.
\newblock Journal of Elasticity. 2024;156:549--569.
\newblock Available from: \url{https://doi.org/10.1007/s10659-024-10062-z}.

\bibitem{tsai1968}
Tsai SW, Pagano NJ.
\newblock Invariant properties of composite materials.
\newblock In: Tsai SW, Halpin JC, Pagano NJ, editors. Composite Materials
  Workshop. Stamford, CT: Technomic; 1968. .

\bibitem{vannucci23c}
Vannucci P.
\newblock Bounds of the technical constants for two-dimensional anisotropic
  elasticity.
\newblock Proceedings of the Royal Society A. 2023;479(20230662).
\newblock Available from: \url{https://doi.org/10.1098/rspa.2023.066}.

\bibitem{werren53}
Werren F, Norris CB.
\newblock Mechanical properties of a laminate designed to be isotropic.
\newblock Madison, WI: US Forest Products Laboratory; 1953. 1841.

\bibitem{CompStruct02}
Vannucci P, Verchery G.
\newblock A new method for generating fully isotropic laminates.
\newblock Composite Structures. 2002;58:75--82.

\bibitem{horio51}
Horio M, Onogi S.
\newblock Dynamic measurements of physical properties of pulp and paper by
  audiofrequency sound.
\newblock Journal of Applied Physics. 1951;22:971--977.

\bibitem{campbell61}
Campbell JG.
\newblock The in-plane elastic constants of paper.
\newblock Australian Journal of Applied Science. 1961;12:356--357.

\bibitem{ostoja2000}
Ostoja-Starzewski M, Stahl DC.
\newblock Random fiber networks and special elastic orthotropy of paper.
\newblock Journal of Elasticity. 2000;60:131--149.

\bibitem{Miki82}
Miki M.
\newblock Material design of composite laminates with required in-plane elastic
  properties.
\newblock In: Proc. of ICCM 4 - Fourth International Conference on Composite
  Materials. Tokio, Japan; 1982. p. 1725--1731.

\end{thebibliography}

\end{document}